\documentclass[letter,12pt]{article}
\usepackage[left=1in, top=1in, bottom=1in, right=1in]{geometry}

\usepackage{latexsym,amssymb,amsmath,color}
\usepackage{graphicx}
\usepackage{cite}
\usepackage{verbatim}
\usepackage[table]{xcolor}

\newcommand{\be}{\begin{equation}}
\newcommand{\ee}{\end{equation}}
\newcommand{\bea}{\begin{eqnarray}}
\newcommand{\eea}{\end{eqnarray}}

\def\1{{\rm 1}}

\def\s1{^{\rm (1)}}

\def\1{{\rm 1}}

\def\s1{^{\rm (1)}}

\usepackage[table]{xcolor}
\usepackage{relsize}
\usepackage{bm}
\usepackage{mathrsfs}
\usepackage{textcomp} % for inverted comma after Sobol
\newcommand{\angstrom}{\mbox{\normalfont\AA}}

\usepackage[latin1]{inputenc}
\usepackage{tikz}
\usetikzlibrary{shapes,arrows}
\tikzstyle{decision} = [diamond, draw, fill=blue!20, 
   text width=5.0em, text badly centered, node distance=3cm, inner sep=0pt]
\tikzstyle{block} = [rectangle, draw, fill=cyan!20, 
   text width=9.0em, text centered, rounded corners]
\tikzstyle{line} = [draw, -latex']
\tikzstyle{cloud} = [draw, ellipse,fill=red!20, node distance=3cm,
   minimum height=2em]
\usepackage{algorithm,algpseudocode,float}
\usepackage{lipsum}
\makeatletter
\newenvironment{breakablealgorithm}
  {% \begin{breakablealgorithm}
   \begin{center}
     \refstepcounter{algorithm}% New algorithm
     \hrule height.8pt depth0pt \kern2pt% \@fs@pre for \@fs@ruled
     \renewcommand{\caption}[2][\relax]{% Make a new \caption
       {\raggedright\textbf{\ALG@name~\thealgorithm} ##2\par}%
       \ifx\relax##1\relax % #1 is \relax
         \addcontentsline{loa}{algorithm}{\protect\numberline{\thealgorithm}##2}%
       \else % #1 is not \relax
         \addcontentsline{loa}{algorithm}{\protect\numberline{\thealgorithm}##1}%
       \fi
       \kern2pt\hrule\kern2pt
     }
  }{% \end{breakablealgorithm}
     \kern2pt\hrule\relax% \@fs@post for \@fs@ruled
   \end{center}
  }
\makeatother
\algdef{SE}[DOWHILE]{Do}{doWhile}{\algorithmicdo}[1]{\algorithmicwhile\ #1}%
\begin{document}

\thispagestyle{empty}
\begin{center}
\textsc{
Uncertainty Quantification in Non-Equilibrium Molecular Dynamics Simulations of
Thermal Transport
}

\bigskip 
\bigskip 

Manav Vohra$^{1}$, Ali Yousefzadi Nobakht$^{2}$, Seungha Shin$^{2}$, Sankaran Mahadevan$^{1}$

\bigskip
\bigskip

\normalsize
$^1$Department of Civil and Environmental Engineering\\
Vanderbilt University\\
Nashville, TN 37235\\

\bigskip

$^2$Department of Mechanical, Aerospace, and Biomedical Engineering\\
The University of Tennessee\\
Knoxville, TN 37996\\

\bigskip

\end{center}

%\vspace{6cm}
%
%\begin{tabbing}
%Corresponding Author: \hspace{5mm} \= Sankaran Mahadevan\\
%       \>  Department of Civil and Environmental Engineering\\
%       \>  Vanderbilt University\\
%       \>  272 Jacobs Hall, VU Mailbox: PMB 351831 \\
%       \>  Nashville, TN 37235 \\
%       \> \\
%Phone: \> (615) 322-3040 \\
%Fax:   \> (615) 343-3773 \\
%Email: \>  sankaran.mahadevan@vanderbilt.edu   \\
%\\
%Submitted to: \> \textit{International Journal of Heat and Mass Transfer} \\
%\>  April 2018\\
%
%\bigskip
%\end{tabbing}
%
%\clearpage

\baselineskip=22pt

%\tableofcontents
%\clearpage

\section*{Abstract}

Bulk thermal conductivity estimates based on predictions from non-equilibrium molecular dynamics (NEMD)
using the so-called direct method are known to be severely under-predicted since finite simulation
length-scales are unable to mimic bulk transport. Moreover, subjecting the system to a temperature gradient
by means of thermostatting tends to impact phonon transport adversely.  Additionally, NEMD predictions
are tightly coupled with the choice of the inter-atomic potential and the underlying values associated with its
parameters. In the case of silicon (Si), nominal estimates of the Stillinger-Weber (SW) potential parameters are largely based 
on a constrained regression approach aimed at agreement with experimental data while ensuring structural
stability. However, this approach has its shortcomings and it may not be ideal to use the same set of parameters
 to study a wide variety of Si-based
systems subjected to different thermodynamic conditions. 
In this study, NEMD simulations are performed on a Si bar to investigate the impact of bar-length,
and the applied thermal gradient on the discrepancy between predictions and the available measurement 
for bulk thermal conductivity at 300~K by constructing statistical response surfaces at different temperatures. 
The approach helps quantify the discrepancy, observed to be largely dependent on the system-size, with minimal
computational effort. A computationally efficient approach based on derivative-based sensitivity measures to
construct a reduced-order polynomial chaos surrogate for NEMD predictions is also presented. The surrogate
is used to perform parametric sensitivity analysis, forward propagation of the uncertainty, and calibration of the important SW potential parameters in a 
Bayesian setting. It is found that only two (out of seven) parameters contribute significantly to the uncertainty
in bulk thermal conductivity estimates for Si. 
%\clearpage

\section{Introduction}
\label{sec:intro}

\begin{comment}
1. Background on use of MD simulations for thermal transport, preferred for studying
thermal transport by phononic interactions (refer notes from book suggested by Amuthan)

2. One approach to NEMD is the Direct Method, commonly used for estimating the bulk
thermal conductivity. A brief discussion on the direct method and associated pros and cons
(notes from Dellan's paper and book suggested by Amuthan) 
Predictions impacted by the choice of potential, values of
individual parameters, size, and potentially due to duration and applied temperature gradients
(cite Amuthan book, Francesco's paper, McGaughey's paper). 
Errors are introduced by thermostatting (Amuthan book). Nominal value of SW potential parameters
based on fitting against experiments and to ensure structural stability etc. (SW paper)

3. Motivate uncertainty analysis and briefly discuss and cite recent efforts (Francesco, Kirby,
Murthy). Highlight focus and key contributions of the present work and how it differs from
those efforts. 

4. Section-wise overview of the paper.  
\end{comment}

Classical molecular dynamics (MD) is commonly used to study thermal transport by means of
phonons in material systems comprising non-metallic elements such
as carbon, silicon, and germanium~\cite{Dumitrica:2010}. 
A major objective for many such studies is the
estimation of bulk thermal conductivity of the system. One of the most commonly used approaches,
regarded as the direct method~\cite{Schelling:2002,Turney:2009,Zhou:2009,Landry:2009,
McGaughey:2006,Ni:2009,Shi:2009,Wang:2009,Papanikolaou:2008}, is a non-equilibrium
technique that involves the application
of a heat flux or a temperature gradient by means of thermostatting, across the system. 
The corresponding steady-state temperature gradient in the former or the heat exchange between
the two thermostats in the latter, is used to estimate the bulk thermal conductivity (at a given
length or size) using 
Fourier's law. However, when the simulation domain is comparable to or smaller than the
mean free path, thermal conductivity estimates from the direct method depends on the
distance between the two thermostats, due to significant contribution of boundary scattering.
Hence, to estimate the bulk thermal conductivity, computations are performed for a range of
system lengths and the inverse
of thermal conductivity is plotted against the inverse of length. The $y$-intercept of a
straight line fit to the observed trend is considered as the bulk thermal conductivity 
estimate. 

Although widely used, the direct method is known to severely under-predict
the bulk thermal conductivity compared to experimental 
measurements~\cite{Haynes:2014,Shanks:1963}. This is primarily
due to length scales used in the simulation that are several orders of magnitude smaller
than those used in an experiment. As a result, the sample length is much smaller than the
bulk phonon mean free path leading to the so-called ballistic transport of the phonons.
The mean free path of such phonon modes is limited to the system size that reduces their
contribution to thermal transport. Moreover, the
introduction of thermostats typically reduces the correlation between vibrations of 
different atoms potentially reducing the thermal conductivity
further~\cite{Evans:2008}. The average temperature gradient experienced by the 
system could thus be different from the simulation input and is a potential source of 
uncertainty. Estimation of thermal conductivity using the
direct method is therefore impacted by the choice of system size and potentially due to
fluctuations in the temperature gradient experienced by the system due to thermostatting.  

Predictions of non-equilibrium molecular dynamics (NEMD) simulations are also dependent on the
choice of the inter-atomic potential as well as values associated with individual parameters
of a given potential. For instance, in the case of crystalline Si, the Stillinger-Weber (SW)
inter-atomic potential is widely used. However, as discussed by Stillinger and Weber
in~\cite{Stillinger:1985}, their proposed nominal values of individual parameters were based
on a limited search in a 7D parameter space while ensuring structural stability and
reasonable agreement with experiments. It is therefore likely that these nominal estimates
for individual parameter values in the SW potential may not yield accurate results for a
wide variety of Si-based systems and warrant further investigation to study the impact of
underlying uncertainties on MD predictions. Along these lines, 
a recent study by Wang et al. performed uncertainty quantification of thermal conductivities
from equilibrium molecular dynamics simulations~\cite{Wang:2017}.
Rizzi et al. focused on the effect of
uncertainties associated with the force-field parameters on bulk water properties using MD
simulations~\cite{Rizzi:2012}. Marepalli et al. in~\cite{Marepalli:2014} considered a stochastic model
for thermal conductivity to account for inherent noise in MD simulations, and study its impact on
spatial temperature 
distribution during heat conduction. Jacobson et al. in~\cite{Jacobson:2014} implemented an uncertainty
quantification framework to optimize a coarse-grained model for predicting the properties of monoatomic
 water. While these are significant contributions, it is only recently 
that researchers have started accounting for the presence of uncertainties in MD predictions in a
systematic manner. There is a definite need for additional efforts aimed at efficiency and accuracy
to enable uncertainty analysis in MD simulations for a wide range of applications.

In the present work, we focus our efforts on uncertainty analysis in the predictions of NEMD simulations
for phonon transport using a silicon bar. An overview of the set-up for the simulations is
provided in Section~\ref{sec:setup}. As discussed earlier, predictions from NEMD
exhibit large discrepancies with experimental observations depending upon system size and potentially
due to fluctuations in the applied temperature gradient. Additionally, the thermal conductivity estimates are
tightly coupled with parameter values associated with the inter-atomic potential. Hence, we set out to
accomplish multiple objectives through this research effort: First, we construct response surfaces
in order to characterize the dependence of discrepancy in thermal conductivity estimates (between
MD simulations and experiments) on system size, and applied temperature gradient~(Section~\ref{sec:response}).
Second, we perform sensitivity analysis to study the impact of SW potential parameter values on 
uncertainty in the predictions~(Section~\ref{sec:sense}).
Third, we exploit our findings from sensitivity analysis to 
construct a reduced order surrogate for uncertainty analysis~(Section~\ref{sec:ros}). 
Fourth, we illustrate the calibration
of important parameters in a Bayesian setting to evaluate their posterior
distributions~(Section~\ref{sec:bayes}). Construction of the
response surfaces, parametric sensitivity analysis, and Bayesian calibration can all be computationally
challenging endeavors especially in situations involving compute-intensive simulations as in the
case of NEMD. We therefore employ polynomial chaos (PC) surrogates~\cite{Xiu:2002,Ghanem:1990}
using non-intrusive spectral approaches~\cite{Olivier:2010} to reduce the computational effort pertaining
to the aforementioned objectives. Moreover, since the construction of a PC surrogate itself can be
expensive, we demonstrate a novel approach in Section~\ref{sec:sense} that implements derivative-based
sensitivity measures~\cite{Sobol:2010} to reduce the dimensionality of the surrogate a priori while
ensuring reasonable predictive accuracy. 

\bigskip
\bigskip
\section{NEMD Set-up}
\label{sec:setup}

\begin{figure}[htbp]
\begin{center}
\begin{tabular}{cc}
  \includegraphics[width=0.48\textwidth]{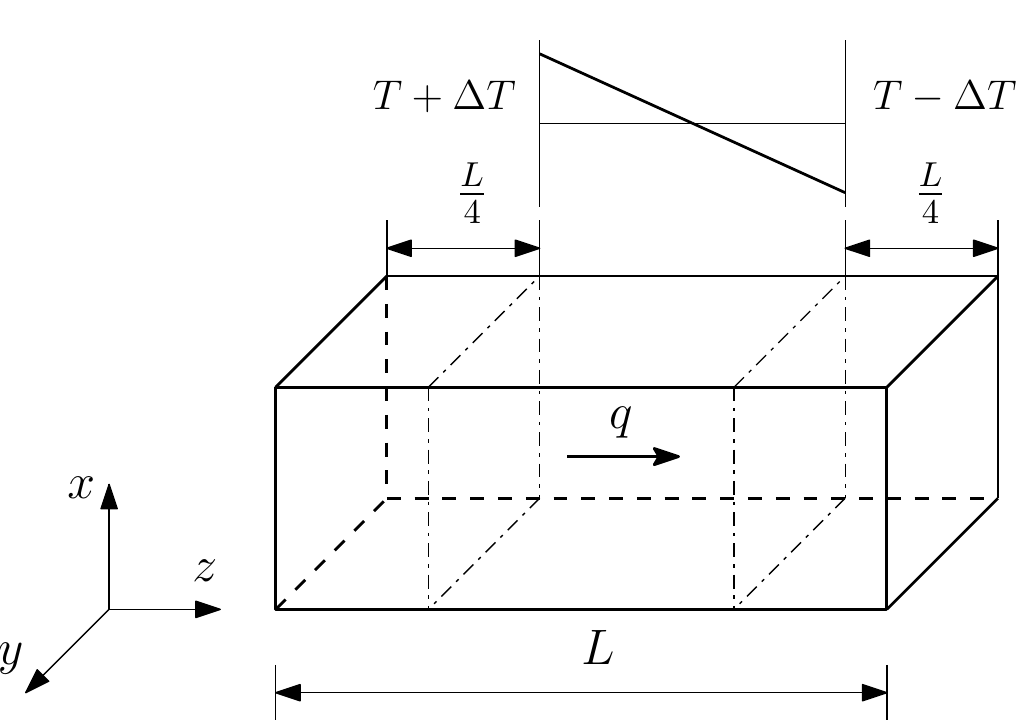}
  &
  \hspace{3mm}
  \includegraphics[width=0.40\textwidth]{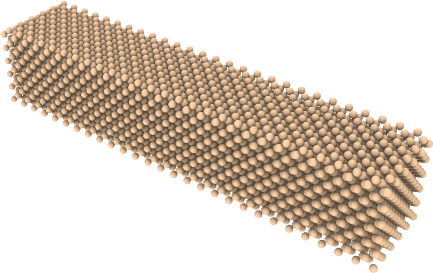}
  \\ (a) & (b)
  \end{tabular}
\caption{(a) Schematic illustration of the set-up for evaluating thermal conductivity of Si using NEMD. (b) 
Arrangement of Si atoms prior to the application of temperature gradient.}
\label{fig:setup}
\end{center}
\end{figure}

In this work, non-equilibrium molecular dynamics (NEMD) simulations are performed using the 
LAMMPS~\cite{Plimpton:2007} software package.
Essentially, a temperature gradient is applied by means of Langevin thermostats located
at $L$/4 and $3L$/4 in a silicon bar of length, $L$ 
as shown using a schematic in Figure~\ref{fig:setup}. 

The set of
inputs to LAMMPS is provided below in Table~\ref{tab:input}. Note that specific values for the length of the bar
and the applied temperature gradient are not provided since we investigate thermal conductivity trends for a range of 
values of the two parameters as discussed later in Section~\ref{sec:response}. A careful analysis focused on
minimizing temperature fluctuations during different stages of the simulation was performed to optimize for the
choice of height and width of the bar as well as the duration of the simulation. 

\begin{table}[htbp]
\begin{center}
\begin{tabular}{|c||c|}
\hline
Lattice Constant, $a$ ($\angstrom$) & 5.43 \\ \hline
Width, Height ($\angstrom$) & 22$a$ \\ \hline
$\Delta t$  (ps) & 0.0005 \\ \hline
Simulation Run Length (ps) & 320 \\ \hline
Boundary Condition & Periodic \\ \hline
Lattice Structure & Diamond \\ \hline
Inter-atomic Potential & Stillinger-Weber \\ 
\hline
\end{tabular}
\end{center}
\label{tab:input}
\caption{Set of inputs for the NEMD simulation to estimate thermal conductivity in a Si bar using LAMMPS.}
\end{table}

The NEMD simulation has three stages associated with it as illustrated below in the flow diagram. In the
first stage, the NVT ensemble equilibrates the system to a specified bulk temperature, i.e., the temperature
at which thermal conductivity is to be estimated. In the second stage, the NVE ensemble equilibrates the
thermostats at their respective temperatures. It is followed by another NVE ensemble that captures the
trajectory of individual atoms and results in a steady state estimate for the thermal energy exchange between
the two thermostats. 

\begin{center}

NVT \hspace{5mm} $\rightarrow$ \hspace{5mm} NVE \hspace{5mm}
$\rightarrow$ \hspace{5mm} NVE
\\ \vspace{1mm}
\tiny \hspace{-5mm}[Equilibrate system to 300 K] \hspace{1mm} [Equilibrate thermostats] \hspace{4mm}
 [Generate Data]
\\ \vspace{1mm}

\tiny{N: Number of Atoms~~~V: Volume~~~T: Temperature~~~E: Energy}
\end{center}

The steady state exchange energy ($q$) is used in Fourier's law to estimate bulk thermal conductivity ($\kappa$)
for the Si  bar:

\be
 \kappa = \frac{q''}{\left|\frac{dT}{dz}\right|} 
\ee

\noindent where $q''$ denotes the steady state heat flux (W/m$^2$) and
$\left|\frac{dT}{dz}\right|$ denotes the magnitude of the applied 
temperature gradient along the direction of heat flow (see~Figure~\ref{fig:setup}(a)).

\bigskip
\bigskip
\section{Response Surface of the Discrepancy}
\label{sec:response}

\begin{figure}[htbp]
 \begin{center}
  \includegraphics[width=0.70\textwidth]{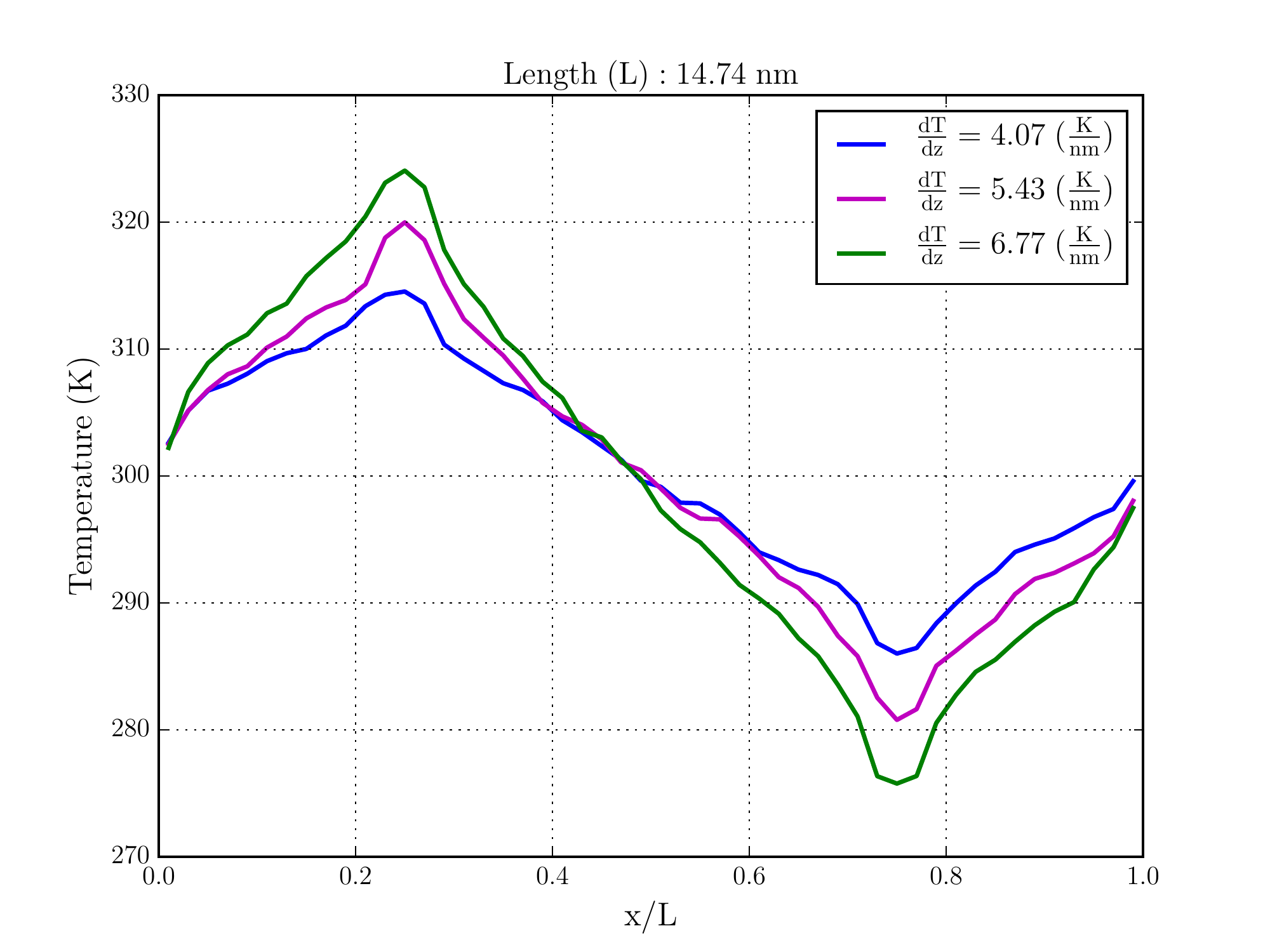}
\caption{Temperature distribution along a Si bar of length 14.74~nm for
different scenarios of applied temperature gradient.}
\label{fig:kapitza}
\end{center}
\end{figure}

As discussed earlier in Section~\ref{sec:intro}, bulk thermal conductivity estimates using NEMD simulations
are lower than measured values primarily due to reduction in mean free path associated with phonon transport. 
Additionally, the introduction of thermostats causes significant fluctuations in the applied temperature gradient,
especially in their vicinity. 
We illustrate this phenomenon by plotting the temperature 
distribution along the length of the bar in Figure~\ref{fig:kapitza}. 

In this section, we focus on the impact of
system size, specifically the length of the Si bar as well as the applied temperature gradient on the discrepancy
in bulk thermal conductivity between NEMD predictions and experimental data. For this purpose, we consider
a range of values for the bar length and the temperature gradient. In order to determine the discrepancy trends, one
might consider evaluating the thermal conductivity using NEMD simulations for a large set of values of length and
temperature gradient. However, considering the computational expense associated with each pair of values, this approach quickly
becomes computationally prohibitive. Instead, we construct a response surface using a 2D 
PC representation of the discrepancy which requires NEMD predictions for a small number of
combinations of length and temperature gradient values as discussed in the following section. 

\subsection{Polynomial Chaos Response Surface}

The PC response surface approximates the functional relationship between independent
and uncertain 
inputs~$(\bm{\theta})$ to a model with the output $\mathcal{Y}$. Essentially, it is a truncated expansion with polynomial 
basis functions that converges in a least-squares sense. For an accurate PC representation, the output should
vary smoothly with respect to the uncertain inputs~\cite{Vohra:2014}  and must
be L-2 integrable:

\be
\mathbb{E}[\mathcal{Y}^2] = \int_{\mathcal{D}_{\bm{\theta}}} \mathcal{Y}^2 \mathbb{P}(\bm{\theta}) 
d\bm{\theta} < \infty
\ee

\noindent where $\mathcal{D}_{\bm{\theta}}$ is the domain of the input parameter space and 
$\mathbb{P}(\bm{\theta})$ is the joint probability distribution of individual components of $\bm{\theta}$.
In the present setting, $\bm{\theta}$:~$\{L,\frac{dT}{dz}\}$ and the output, $\mathcal{Y}$ is the
discrepancy~($\epsilon_{\mbox{\tiny{d}}}$ = 
$\lvert\kappa_{\mbox{\tiny{MD}}}$ - $\kappa_{\mbox{\tiny{E}}}\rvert$)
in bulk thermal conductivity predictions from 
NEMD~($\kappa_{\mbox{\tiny{MD}}}$) and experimental data~($\kappa_{\mbox{\tiny{E}}}$), at a 
given temperature, $T$. The PC representation of $\epsilon_{\mbox{\tiny{d}}}$ is given as:

\be
\epsilon_{\mbox{\tiny{d}}} \approx \mathcal{\epsilon}_{\mbox{\tiny{d}}}^{\mbox{\tiny{PCE}}} = 
\sum_{\bm{k}\in\mathcal{I}} c_{\bm{k}}(T)\Psi_{\bm{k}}(\bm{\xi(\theta)}) 
\ee

\noindent Individual components of the uncertain input vector, $\bm{\theta}$ are parameterized in terms of canonical random 
variables, $\bm{\xi}$ distributed uniformly in the interval $[-1,1]$. 
 $\Psi_{\bm{k}}$'s are multivariate polynomial basis functions, orthonormal with respect to the joint probability 
 distribution of $\bm{\xi}$. The degree of truncation in the above expansion is denoted by $\bm{k}$, a subset of
 the multi-index set $\mathcal{I}$ that comprises of individual degrees of univariate polynomials in $\Psi_{\bm{k}}$.
The PC coefficients, $c_{\bm{k}}$'s can be estimated using either numerical quadrature or advanced techniques
involving basis pursuit de-noising~\cite{Peng:2014}, compressive
sampling~\cite{Hampton:2015}, and least angle regression~\cite{Blatman:2011} suited for large-dimensional
applications. However, in our case, since the response surface is 2D, we use Gauss-Legendre quadrature to
obtain accurate estimates of the PC coefficients. 
\bigskip

In order to construct the response surface of the discrepancy, we consider respective intervals for the Si bar length,
$L$ and the applied temperature gradient, $\frac{dT}{dz}$ as $[50a,100a]$~($\angstrom$) and
 $[\frac{1.5}{a},\frac{2.5}{a}]$~($\frac{\mbox{\tiny{K}}}{\tiny{\angstrom}}$); $a$ being the lattice constant. For the considered length interval, the number of atoms in the
simulation were observed to range from 201344 to 379456. It must be noted that our 
focus is on illustrating a methodology for constructing a response surface that
sufficiently captures the relationship between the inputs: $L$, $\frac{dT}{dz}$ and
the discrepancy, $\epsilon_d$. Hence, we believe
that the considered range for $L$ is large enough to be able to capture the input-output
trends as presented further below, while ensuring reasonable computational effort.
Additionally, the underlying noise associated with NEMD simulations was found to be
negligible in the considered range for $\frac{dT}{dz}$ since we use average temperature
at the bin point during the data production NVE ensemble.
 
\begin{figure}[htbp]
\begin{center}
\begin{tabular}{cc}
 \hspace{-10mm}
  \includegraphics[width=0.55\textwidth]{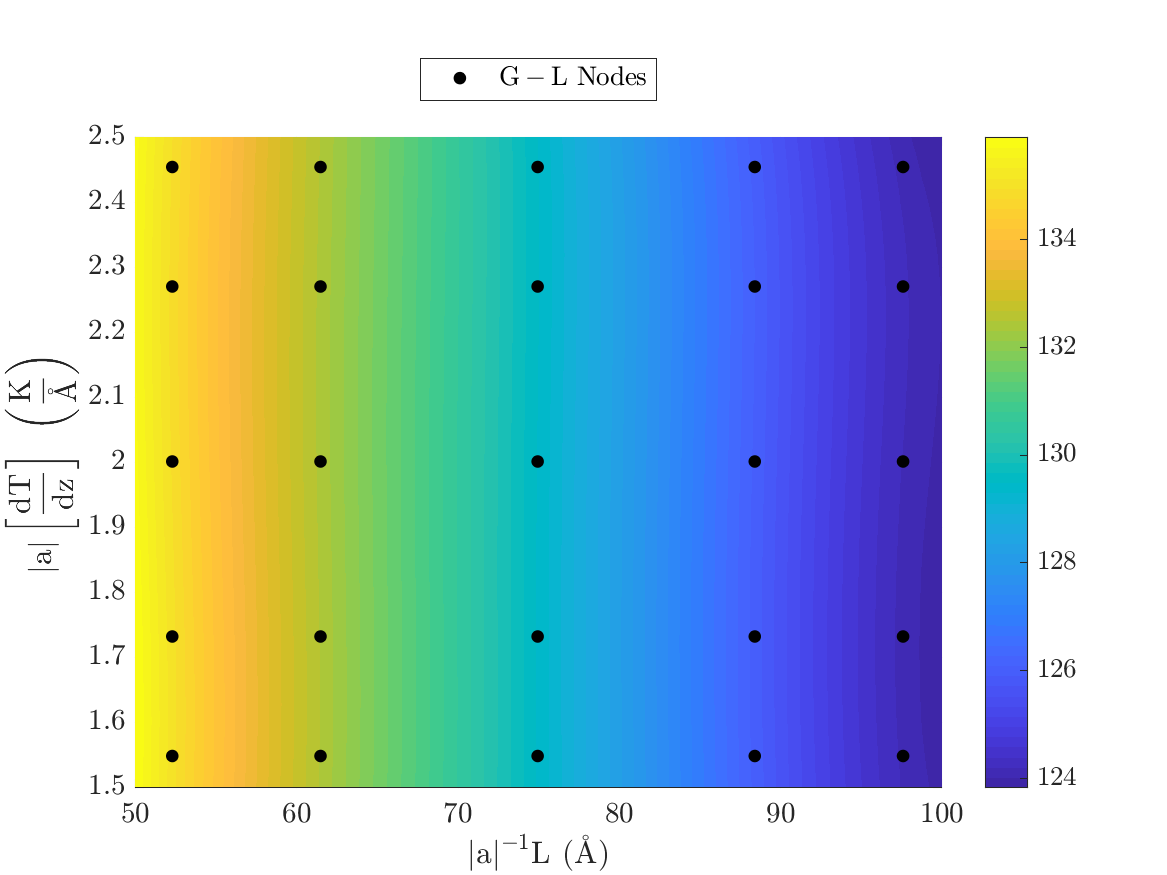}
  &
  %\hspace{-9mm}
  \includegraphics[width=0.55\textwidth]{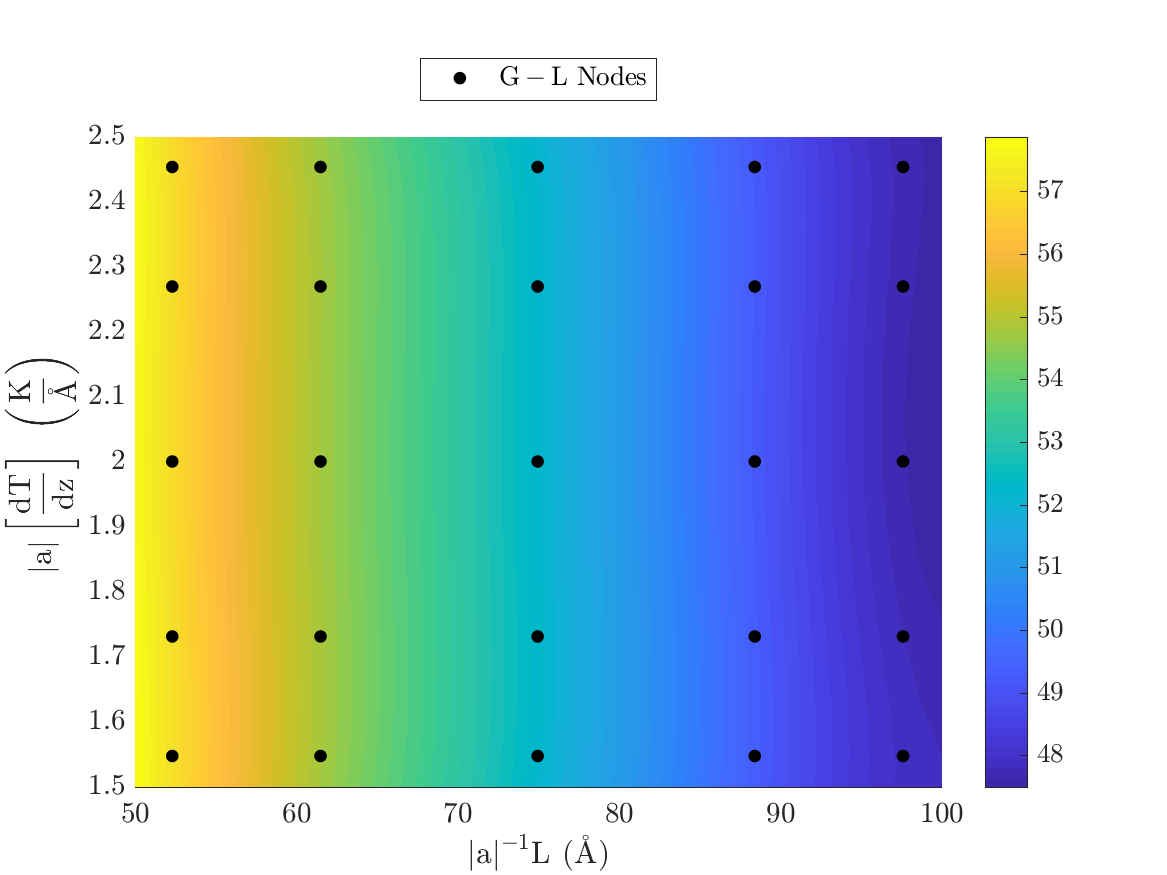}
  \\ (a) & (b)
  \end{tabular}
  \\ \vspace{1mm}
  \includegraphics[width=0.50\textwidth]{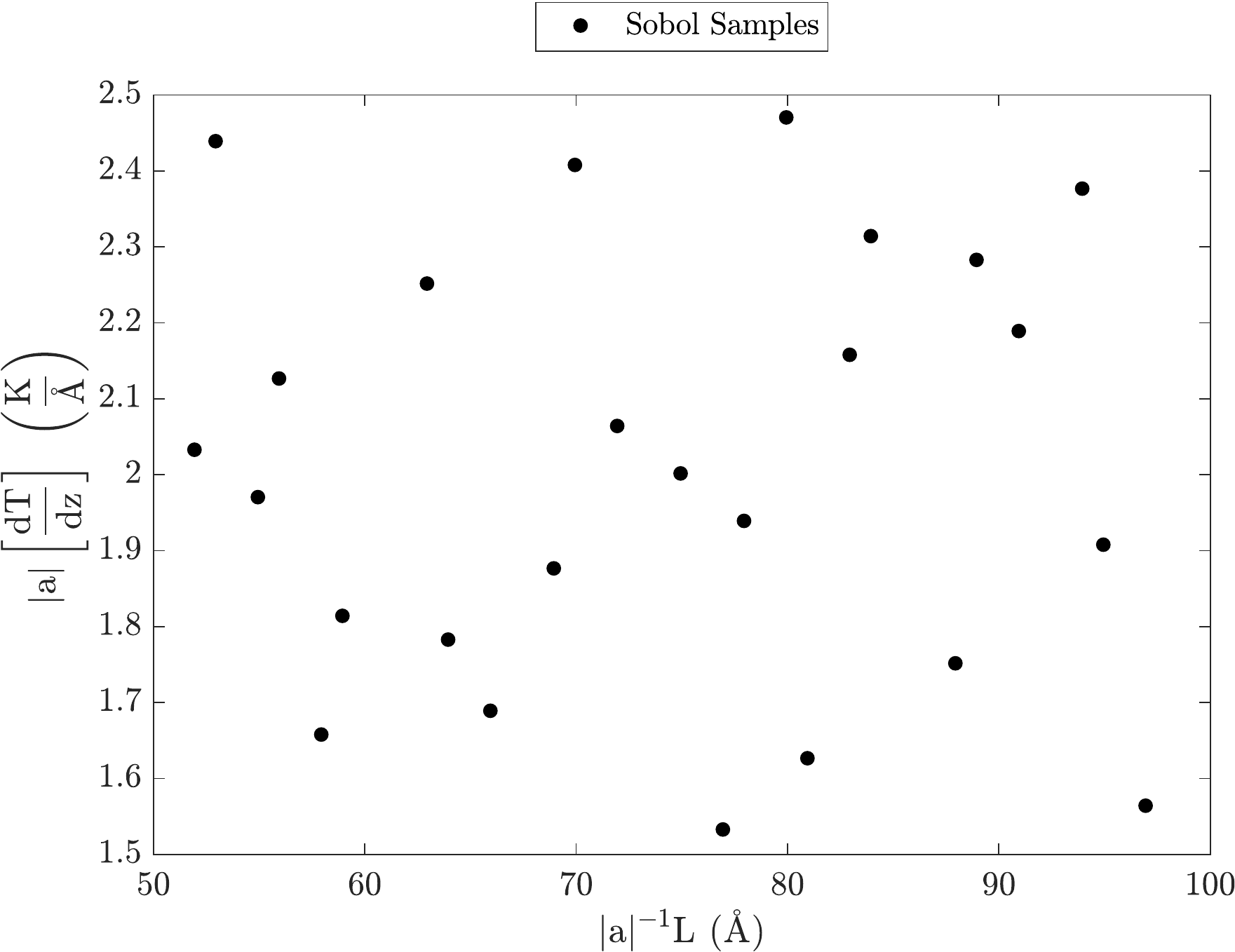}
  \\ (c)
\caption{Response surface of the discrepancy in bulk thermal conductivity at (a) $T$ = 300~K and
 (b) $T$ = 500~K. Gauss-Legendre quadrature nodes are highlighted in both cases. (c) Sobol samples
 in the 2D domain used for verifying the accuracy of the response surfaces. Here, $|a|$ denotes the magnitude
 of the lattice constant.}
\label{fig:rs2}
\end{center}
\end{figure}

NEMD predictions of $\kappa_{\text{MD}}$
are obtained at 25 Gauss-Legendre quadrature nodes (or points) in the 2D parameter 
space as highlighted in Figure~\ref{fig:rs2}(a)~and~\ref{fig:rs2}(b). 
Note that 5 points are considered along $L$ and
$\frac{dT}{dz}$ assuming that a fourth order polynomial basis would be sufficient in
both cases, and the associated computational effort is not too large. 
Later, we assess the validity of our assumption by verifying the accuracy of the resulting
response surface.
 
\begin{figure}[htbp]
\begin{center}
\begin{tabular}{cc}
  \hspace{-12mm}
  \includegraphics[width=0.60\textwidth]{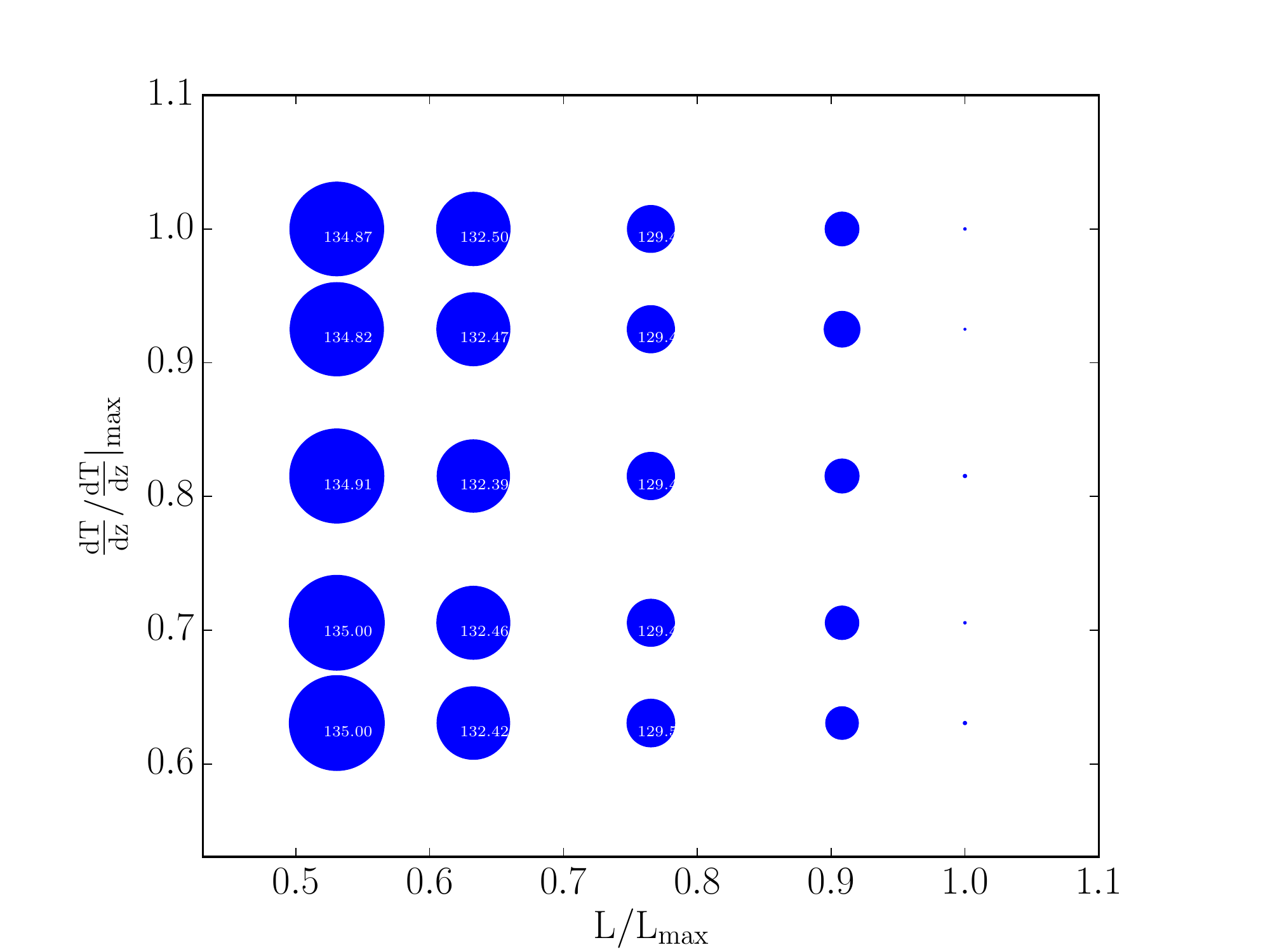}
  &
  \hspace{-9mm}
  \includegraphics[width=0.56\textwidth]{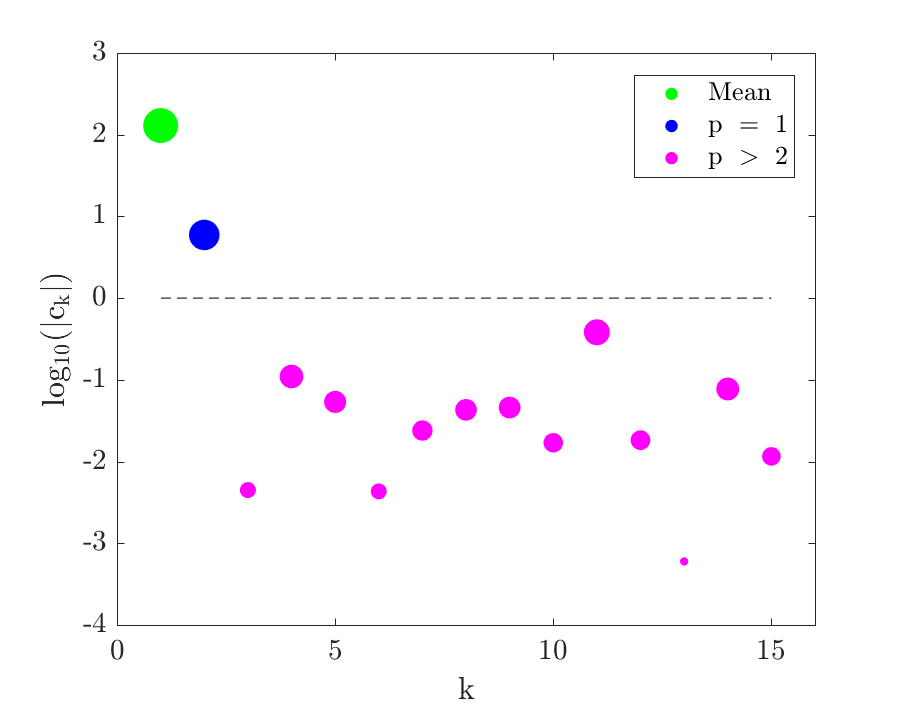}
  \\ (a) & (b)
  \end{tabular}
\caption{(a) Realizations of discrepancy in bulk thermal conductivity at the Gauss-Legendre quadrature notes are
depicted using circles. The size of the circle in each case is proportional to the discrepancy estimate, also provided
 in cases where it is observed to be relatively large. (b) Spectrum of PC coefficients is depicted using circles of
 varying sizes, proportional to the log value of their magnitude. The above computations were performed at 300 K.}
\label{fig:rs1}
\end{center}
\end{figure}

 The discrepancy between $\kappa_{\mbox{\tiny{MD}}}$ and $\kappa_{\mbox{\tiny{E}}}$ is computed at the quadrature nodes as illustrated in Figure~\ref{fig:rs1}(a) to estimate the PC coefficients. The spectrum of 
 resulting PC coefficients is illustrated in Figure~\ref{fig:rs1}(b). 
Note that the value of $\kappa_{\mbox{\tiny{E}}}$ is considered to be 149~W/m/K as provided 
 in~\cite{Shanks:1963}. Response surfaces constructed at the bulk temperature, $T$ = 300~K and 500~K are 
 illustrated
 in Figure~\ref{fig:rs2}(a) and~\ref{fig:rs2}(b) respectively.

As expected, the discrepancy is observed to decrease
 with the bar length ($L$) due to increase in the mean free path. It is however interesting to note that the 
 variation in discrepancy due to changes in the applied temperature gradient in the considered range is found to be
 negligible. The accuracy of the response surface is verified by computing a relative L-2 norm of the error
 ($\varepsilon_{\tiny{\mbox{L-2}}}$) on an independent set of Sobol samples~\cite{Saltelli:2010}
(see Figure~\ref{fig:rs2}(c)) in the 2D 
 parameter domain as follows:
 
 \be
 \varepsilon_{\tiny{\mbox{L-2}}} = \frac{\left[\sum_{j}(\epsilon_{\tiny{\mbox{d},j}}^{\mbox{\tiny{MD}}} - 
 \epsilon_{\tiny{\mbox{d},j}}^{\mbox{\tiny{PCE}}})^2\right]^{\frac{1}{2}}}{\left[\sum_j(\epsilon_{\tiny{\mbox{d},j}}
 ^{\mbox{\tiny{MD}}})^2\right]^{\frac{1}{2}}} 
 \label{eq:err_l2}
 \ee
 
\begin{figure}[htbp]
\begin{center}
\begin{tabular}{cc}
 \hspace{-10mm}
  \includegraphics[width=0.50\textwidth]{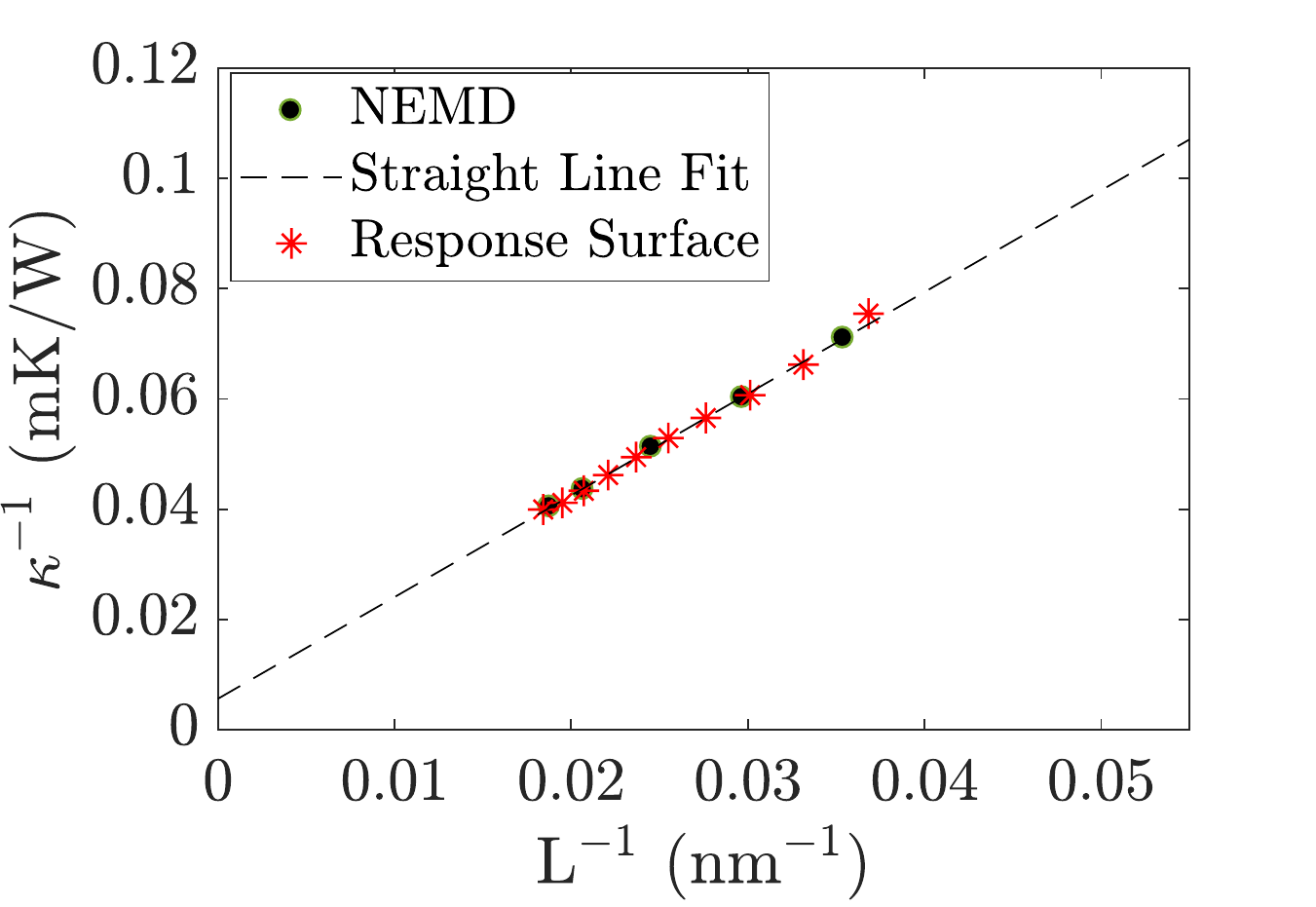}
  &
  %\hspace{-9mm}
  \includegraphics[width=0.50\textwidth]{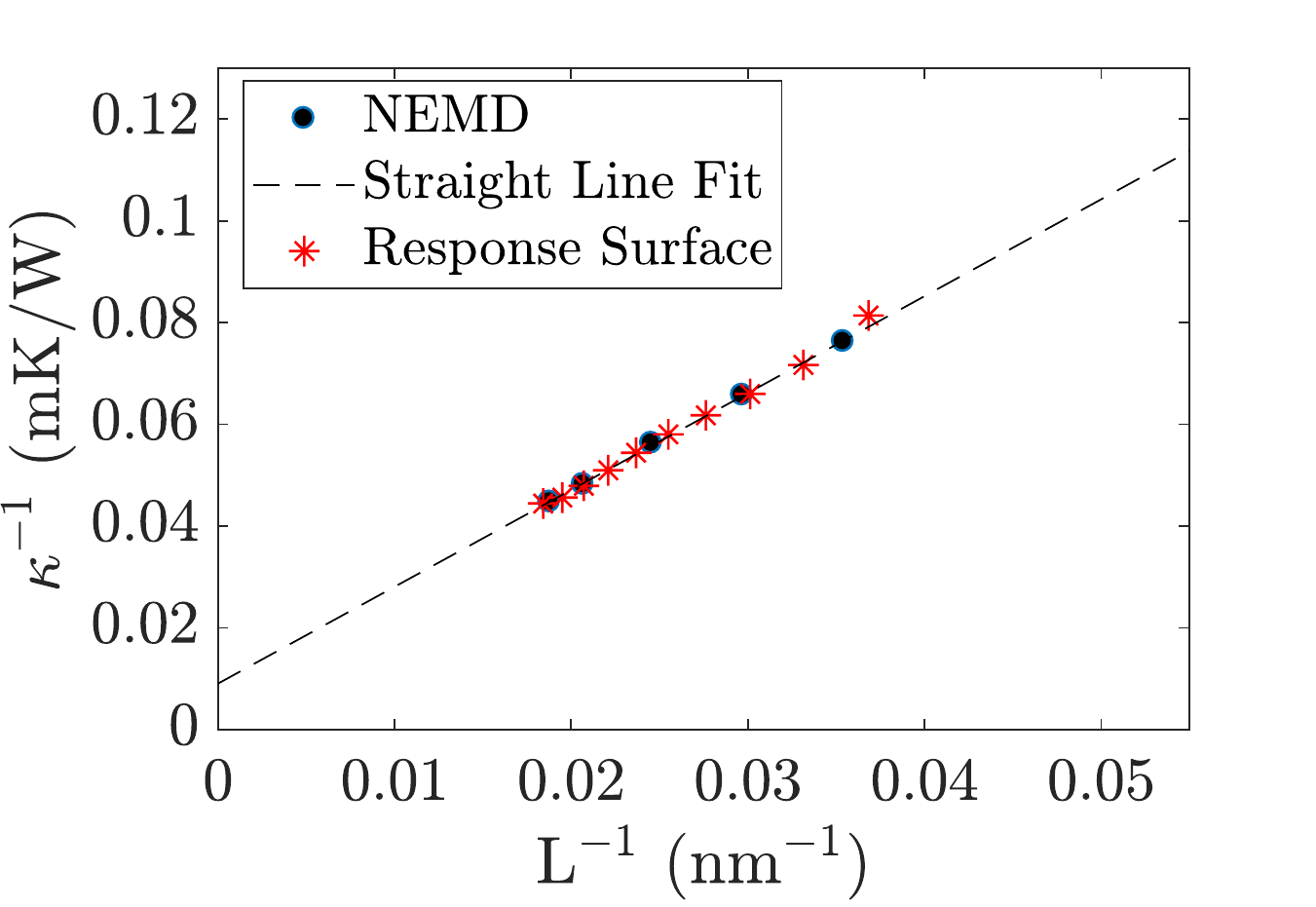}
  \\ (a) & (b)
  \end{tabular}
 \caption{Inverse of the bulk thermal conductivity estimates are plotted against the inverse of Si bar length
 using predictions from NEMD as well as estimates from the response surface at (a)  $T$ = 300~K and 
 (b) $T$ = 500~K. A straight line fit to the estimates is also illustrated in each case.}
\label{fig:kinv}
\end{center}
\end{figure}

 A response surface was also constructed at $T$ = 1000~K (plot not included for brevity) and the impact of
 varying the temperature gradient on the discrepancy was still observed to be negligible. In all cases,
 $\varepsilon_{\tiny{\mbox{L-2}}}$ in Eq.~\ref{eq:err_l2} was estimated to be of $\mathcal{O}(10^{-3})$ thereby
 indicating that the response surfaces could be used to predict the discrepancy for a given
 point ($L$,$\frac{dT}{dz}$) in the considered domain with reasonable accuracy. As an additional verification
 step, we plot the inverse of thermal conductivity against the inverse of bar length using data from NEMD
 simulations as well as predictions from the response surface 
($\kappa_{\mbox{\tiny{MD}}}~=~\kappa_{\mbox{\tiny{E}}}~-~\epsilon_{\mbox{\tiny{d}}}$) in Figure~\ref{fig:kinv}.
It is observed that the response surface estimates exhibit an expected linear trend, consistent with NEMD
  predictions. Using the y-intercept of the straight line fit, the bulk thermal conductivity was estimated
 as 178.52 W/m/K and 110.64 W/m/K at 300~K and 500~K respectively. Using a wider range for system length would
help improve these estimates. However, as discussed earlier, our focus is on demonstrating
a methodology for quantifying the uncertainties (with limited resources) in bulk thermal conductivity predictions
 using NEMD as opposed
to determining accurate estimates for the same. Constructing response surfaces using a
 relatively small number of NEMD predictions thus offers
  potential for huge computational savings for studies aimed at predicting thermal conductivity trends and 
  quantifying the discrepancy with measurements for a wide range of system sizes, applied temperature gradients as 
  well as bulk temperatures.

In the following sections, we shift our focus towards understanding the impact of the uncertainty
in SW potential parameter values on the uncertainty in NEMD predictions for bulk
thermal conductivity in Si.

\bigskip
\bigskip
\section{Sensitivity Analysis of the Inter-atomic Potential}
\label{sec:sense}

As discussed earlier in Section~\ref{sec:intro}, bulk thermal conductivity estimates in NEMD
are dependent on the choice of the inter-atomic potential as well as values associated with
the individual potential parameters. In the case of silicon, the Stillinger-Weber (SW)
is a commonly used inter-atomic potential~(see~\cite{Laradji:1995,Zhang:2014,Jiang:2015,Watanabe:1999,Zhou:2013} and references therein). Functional form of the SW potential is given as follows:
\be
\Phi = \sum\limits_{i,j(i<j)}\phi_2(A,B,p,q,\alpha)\hspace{1mm}+\sum\limits_{i,j,k(i<j<k)}\phi_3(\lambda,\gamma)
\ee
where $\phi_2$ and $\phi_3$ represent two-body and three-body atomic interactions 
respectively~\cite{Stillinger:1985}.

Although the SW potential is used for a wide variety of Si systems, 
according to Stillinger and Weber, the set of nominal
values as provided below in Table~2 were based on a constrained search in the 7D parameter space to ensure structural stability and agreement with the available experimental data~\cite{Stillinger:1985}.

\begin{table}[htbp]
\begin{center}
\begin{tabular}{|c|c|c|c|c|c|c|}
\hline 
$A$ & $B$ & $p$ & $q$ & $\alpha$ & $\lambda$ & $\gamma$ \\
\hline \hline
7.049556277 & 0.6022245584 & 4.0 & 0.0 & 1.80 & 21.0 & 1.20 \\
\hline
\end{tabular}
\end{center}
\caption{Nominal values of the parameters of the Stillinger-Weber inter-atomic
potential~\cite{Stillinger:1985}.}
\end{table}

It is noteworthy that the underlying analysis which led to these estimates of the nominal values did not
account for the presence of uncertainty due to
measurement error, noise inherent in MD predictions, inadequacies pertaining to the potential function,
and parametric uncertainties. It is therefore likely that the proposed nominal estimates could be 
improved depending upon the application. Hence, it is critical to understand the effects of uncertainty in
SW potential parameters on bulk thermal conductivity predictions using NEMD. For this purpose, a possible
approach could involve a global sensitivity analysis of NEMD predictions on the SW potential parameters 
by estimating the so-called Sobol\textquotesingle~indices~\cite{Sobol:2001}. However, obtaining converged estimates of
Sobol\textquotesingle~indices typically requires tens of thousands of model evaluations to be able to numerically approximate
multi-dimensional integrals associated with the expectation and variance operators, especially in case $\bm{\theta}$,
the vector of uncertain model inputs is high-dimensional:

\be
 \mathcal{T}_i = \frac{\mathbb{E}_{\bm{\theta}\sim i}[\mathbb{V}_{\theta_i}(\mathcal{G}|\bm{\theta}_{\sim i})]}{\mathbb{V}(\mathcal{G})} 
 \ee
 
 \noindent where $\mathcal{T}_i$ is the Sobol\textquotesingle~total-effect index, $\mathcal{G}$ denotes the model output,
 $\mathbb{E}_{\bm{\theta}\sim i}[]$ is an expectation over all but the $i^{\mbox{th}}$ component of 
 $\bm{\theta}$, and $\mathbb{V}_{\theta_i}()$ is the variance taken over $\theta_i$. Li and Mahadevan
 recently proposed a computationally efficient method for estimating the first-order Sobol\textquotesingle~index~\cite{Li:2016}.
However, since NEMD is compute-intensive,
estimating the Sobol\textquotesingle~indices directly would be impractical in the present scenario. Hence, instead of 
estimating Sobol\textquotesingle~sensitivity
indices, we focus our attention on the upper bound of the Sobol\textquotesingle~total-effect index to determine the relative
importance of SW potential parameters. 
It is observed that for a given application, it might be possible to 
converge to the upper bound on
Sobol\textquotesingle~index with only a few iterations~($\mathcal{O}(10^{1})$)~\cite{Kucherenko:2016}. 
In that case, estimates of the upper bound could be used in lieu of the
Sobol\textquotesingle~indices to determine relative importance of the parameters and hence 
reduce the associated computational
effort by several order of magnitude. The upper bound of the
Sobol\textquotesingle~total effect index\footnote{Sobol\textquotesingle~total effect index is a measure
 of the contribution of an 
input to the variance of the model output, also accounting for the contribution coupled with other inputs.}
($\mathcal{T}_i$) can be expressed in terms of a derivative-based sensitivity measure~(DGSM), $\mu_i$, the Poincar\' e 
constant~($\mathcal{C}_i$), and the total variance of the observed
quantity~($V$)~\cite{Lamboni:2013,Roustant:2014} as follows:   

\be
\mathcal{T}_i \leq \frac{\mathcal{C}_i\mu_i}{V}~(\propto \hat{\mathcal{C}_i\mu_i}) 
\ee 

\noindent The derivative-based sensitivity measure, $\mu_i$ for a given parameter, $\theta_i$ is
defined as an expectation
of the derivative of the output ($G(\bm{\theta})$) with respect to that parameter:

\be
\mu_i = \mathbb{E}\left[\left(\frac{\partial G(\bm{\theta})}{\partial \theta_i}\right)^{2}\right]
\label{eq:mu}
\ee

\noindent Latin hypercube sampling in the 7D parameter space is used to estimate $\mu_i$. Note that $G$ must 
exhibit a smooth variation with each parameter so that the derivative in Eq.~\ref{eq:mu} can be estimated
with reasonable accuracy, analytically or numerically. 
We define a normalized quantity, $\hat{\mathcal{C}_i\mu_i}$ to ensure that its summation over all parameters is 1:

\be
\hat{\mathcal{C}_i\mu_i} = \frac{\mathcal{C}_i\mu_i}{\sum_i \mathcal{C}_i\mu_i} 
\ee

\noindent The choice of $\mathcal{C}_i$ is specific to the marginal probability distribution of the uncertain model
parameter, $\theta_i$. 
%We consider all uncertain parameters to be uniformly
%distributed in the interval~$[a,b]$ in which case $\mathcal{C}_i$  is given as $(b-a)^{2}/\pi^2$~\cite{Roustant:2014}.
The underlying methodology for implementing DGSM to the present application involving thermal transport in bulk Si,
and our key findings are discussed in the following section. 

\subsection{DGSM for SW potential parameters}
\label{sub:dgsm} 

We aim to compute the DGSM and hence the corresponding upper bound on the
Sobol\textquotesingle~total effect index~($\mathcal{T}_i$) for each parameter in the SW potential. For this purpose, we 
introduce small perturbations~($\mathcal{O}(10^{-5})n_i$; $n_i$ being the nominal value) to the nominal values
associated with each parameter and estimate the partial derivatives in Eq.~\ref{eq:mu} using finite difference. 
Hence, in order to compute $\mu_i$ using $N$ points in the d-dimensional parameter space, we require $N(d+1)$
model realizations. The SW potential parameters are considered to be uniformly distributed in a small interval
around the nominal value in which case $\mathcal{C}_i$ is given as $(u-l)^{2}/\pi^2$~\cite{Roustant:2014}; $u$
and $l$ being the upper and lower bounds of the interval respectively.   

\begin{figure}[htbp]
\begin{center}
\begin{tabular}{cc}
  \includegraphics[width=0.40\textwidth]{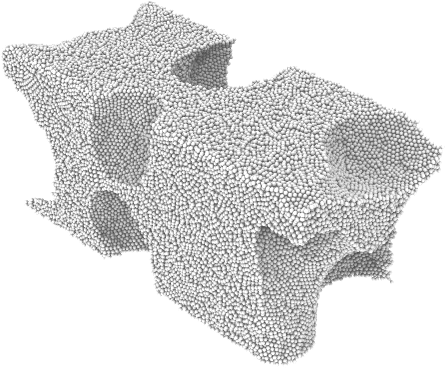}
  &
  \hspace{3mm}
  \includegraphics[width=0.50\textwidth]{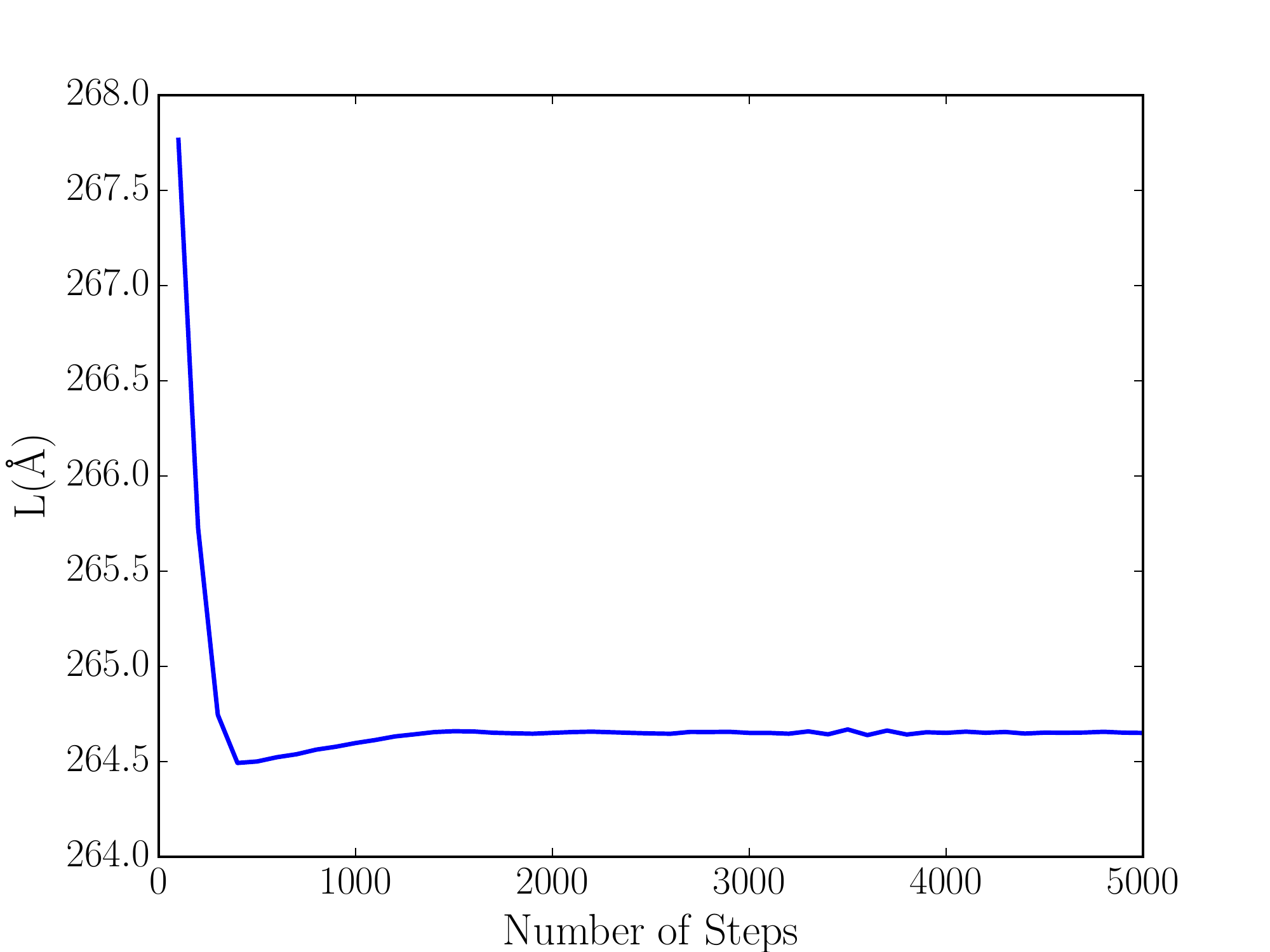}
  \\ (a) & (b)
  \end{tabular}
\caption{(a) A snapshot of the arrangement of atoms illustrating loss of
structural integrity of the Si bar. 
(b) Si bar length is plotted against the number of time steps during
the NPT ensemble stage of the simulation.}
\label{fig:dgsm1}
\end{center}
\end{figure}

Performing NEMD simulations using perturbed values of the SW potential parameters could however be challenging.
For certain combinations of the SW potential parameter values, the steady-state thermal
energy exchange between the thermostats was found to be non-physical at the end of the simulation. We believe that
this was observed in situations 
where the structure had deviated too far from the equilibrium state, and therefore the
structural integrity of the bar was lost as illustrated in Figure~\ref{fig:dgsm1}(a). 

To avoid this issue, we added an NPT ensemble prior to NVT in the NEMD simulation as shown in the
following diagram:

\begin{center}

NPT \hspace{5mm} $\rightarrow$ \hspace{5mm} NVT \hspace{5mm} $\rightarrow$ \hspace{5mm} NVE \hspace{5mm}
$\rightarrow$ \hspace{5mm} NVE
\\ \vspace{1mm}
\tiny [Relax the system]~[Equilibrate system to 300 K] \hspace{1mm} [Equilibrate thermostats] \hspace{4mm}
 [Generate Data]
\\ \vspace{1mm}

\tiny{N: Number of Atoms~~~P: Pressure~~~V: Volume~~~T: Temperature~~~E: Energy}
\end{center}

\noindent The NPT stage of the simulation was allowed to continue for a sufficiently long duration to ensure
that the system is relaxed to a steady value of the bar length as shown in Figure~\ref{fig:dgsm1}(b).

The following algorithm provides the sequence of steps that were used to obtain approximate estimates of the
sensitivity measures for the SW potential parameters. Note that the algorithm has been adapted to the 
specific application in this work. A generalized methodology with a more detailed discussion and its application
to different class of problems will be presented in~\cite{Vohra:2018}.

\bigskip

\begin{breakablealgorithm}
  \caption{Estimating parameter ranks using DGSM.}
  \begin{algorithmic}[1]
    \Procedure{DGSM}{}
      \State Generate $n_1$ points in $\mathbb{R}^{d}$.\Comment{$d$: 
             Number of parameters i.e. 7 in this case}
      \State Perturb each point along the $d$ directions to obtain a set of $n_1(d+1)$ points.
      \State Compute $\mu_i$ using model evaluations at the $n_1(d+1)$ points in Eq.~\ref{eq:mu}
      \State Determine initial ranks, $\mathcal{R}^{old}$ based on $\hat{\mathcal{C}_i\mu_i}$ values for $\theta_i$.
      \State set $k$ = 1\Comment{Iteration counter}
      \Do
        \State Generate $n_k$ new points in $\mathbb{R}^{d}$.
        \State Perturb each point along the $d$ directions to obtain a set of $n_k(d+1)$ points.
        \State Compute and store model evaluations at the $n_k(d+1)$ points.
        \State Compute $\mu_i$ using prior model evaluations at $(d+1)(n_1 + \sum_j^k n_j)$ points.
        \State Determine new ranks, $\mathcal{R}^{new}$ based on updated $\hat{\mathcal{C}_i\mu_i}$ values.
        \State Compute $max\_pdev$ = max$\left(\frac{|\mu_{i,k} - 
               \mu_{i,k-1}|}{ \mu_{i,k-1}}\right)$.\Comment{$max\_pdev$:
               Maximum percentage deviation in $\mu_i$ between successive iterations.}
        \State set $k$ = $k$ + 1
      \doWhile{($\mathcal{R}^{\tiny{new}}$ $\neq$ $\mathcal{R}^{\tiny{old}}$ {\bf or}  
               $max\_pdev$~$>~\tau$)\Comment{$\tau$:~Tolerance}}
    \EndProcedure
  \end{algorithmic}
\end{breakablealgorithm}

\bigskip

%\texttt{Algorithm}
%
%\begin{algorithm}[H]
%\SetAlgoLined
%%\nonl \scriptsize{\textsc{Part I: Parameter Screening}}
%%\nonl \textbf{\texttt{Algorithm}}\;
%\texttt{Generate $n_1$ points in $\mathbb{R}^{d}$}\;
%\texttt{Perturb each point along the $d$ directions to obtain a set of $n_1(d+1)$ points}
%\texttt{\color{blue} $\%$~$d$: Number of parameters in the SW potential i.e. 7}\;
%\texttt{Compute $\mu_i$ using model evaluations at the $n_1(d+1)$ points in Eq.~\ref{eq:mu}}\;
%\texttt{Determine initial ranks, $\mathcal{R}^{old}$ of the parameters based on $\hat{\mathcal{C}_i\mu_i}$ values}\;
%\texttt{set $k$ = 1}
%\texttt{\color{blue}$\%$~$k$:~Iteration counter}\;
%\Repeat{\texttt{($\mathcal{R}^{\tiny{new}}$ $\neq$ $\mathcal{R}^{\tiny{old}}$ $\&$ 
%$max\_pdev$~$>\tau$)~\color{blue}$\%$~$\tau$:~Tolerance}}{
%\texttt{Generate $n_k$ new points in $\mathbb{R}^{d}$}\;
%\texttt{Perturb each point along the $d$ directions to obtain a set of $n_k(d+1)$ points}\;
%\texttt{Compute and store model evaluations at the $n_k(d+1)$ points}\;
%\texttt{Compute $\mu_i$ using prior model evaluations at $(d+1)(n_1 + \sum_j^k n_j)$ points}\;
%\texttt{Determine new ranks, $\mathcal{R}^{new}$ based on updated $\hat{\mathcal{C}_i\mu_i}$ values}\;
%\texttt{Compute $max\_pdev$ = max($\frac{|\mu_{i,k} - \mu_{i,k-1}|}{ \mu_{i,k-1}}$)}
%\texttt{\color{blue}$\%$~$max\_pdev$: Maximum percentage deviation in $\mu_i$ between successive iterations.}\;
%\texttt{set $k$ = $k$ + 1}\;
%}
%\end{algorithm}

\begin{figure}[htbp]
 \begin{center}
  \includegraphics[width=0.5\textwidth]{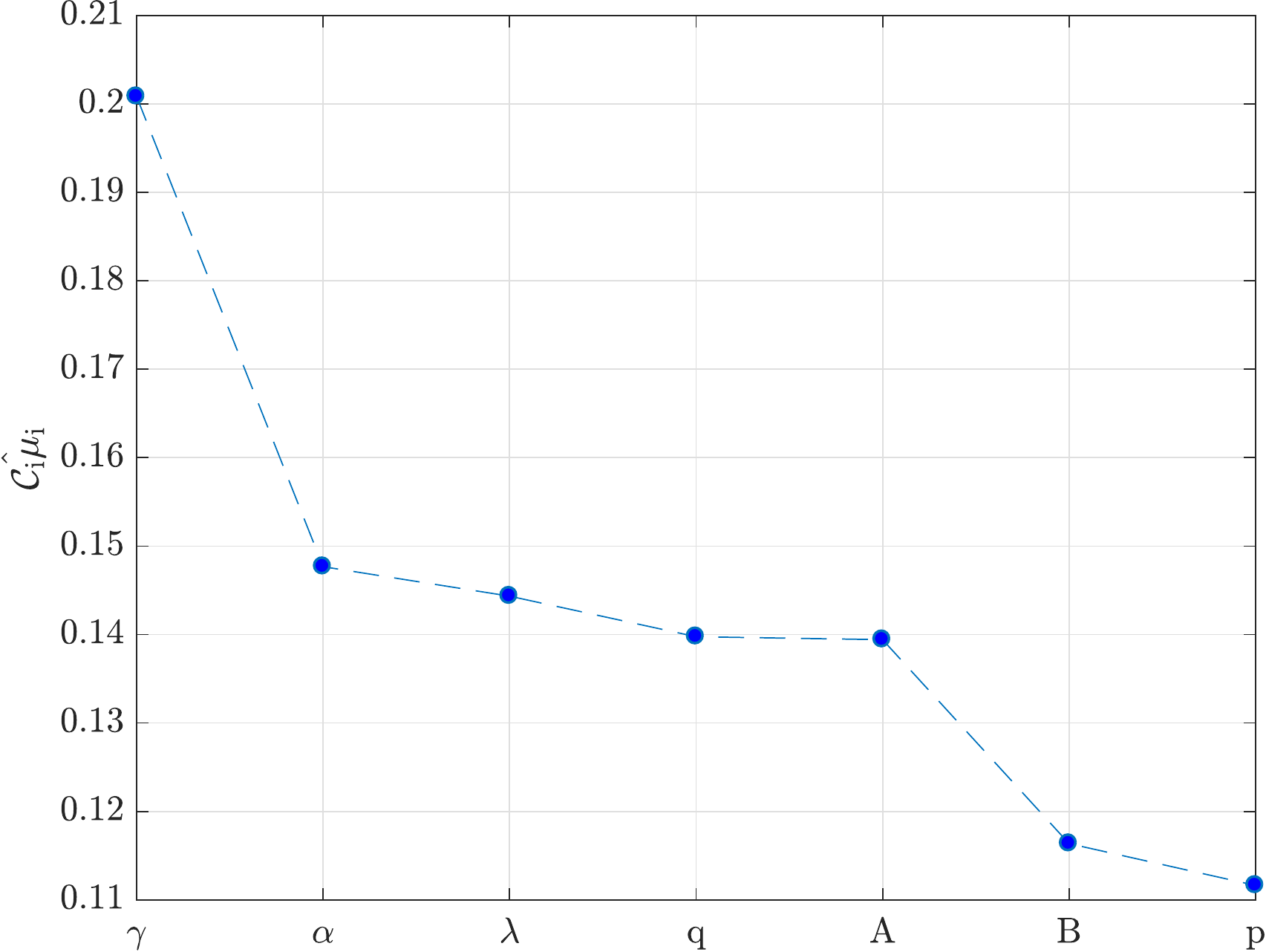}
\caption{The quantity $\hat{\mathcal{C}_i\mu_i}$ as computed after 4 iterations and data at 200
points is plotted for each SW potential parameter.}
\label{fig:ub}
\end{center}
\end{figure}

For the present application, we begin with $n_1$ = 10 samples in the 7D parameter space and add 5 points
at each iteration. Using a tolerance, $\tau$ = 0.05, the above algorithm took 4 iterations i.e. 25 points to 
provide approximate estimates for $\mu_i$. Since finite difference was used to estimate the derivatives in
Eq.~\ref{eq:mu}, it required 25(7+1) i.e. 200 MD runs. It must be noted that although the computational effort
pertaining to the estimation of DGSM can be substantial, it is nevertheless several orders of magnitude smaller
than directly estimating the Sobol\textquotesingle~indices as mentioned earlier. 
In Figure~\ref{fig:ub}, we plot $\hat{\mathcal{C}_i\mu_i}$ as obtained for the SW parameters at the 
end of 4 iterations. It appears that $\gamma$ is significantly more important than other parameters, whereas NEMD
predictions are relatively less sensitive to $B$ and $p$. Large sensitivity towards 
$\gamma$ and $\alpha$ (cut-off radius) is in fact expected since  these two parameters impact the lattice
constant and hence the mean free path associated with the bulk thermal conductivity directly. 

In the following section, we exploit these observations based on DGSM to construct a reduced order surrogate.
The surrogate enables forward propagation of uncertainty in the SW potential to the bulk thermal conductivity 
estimates, and the estimation of Sobol\textquotesingle~sensitivity indices with minimal computational effort.

\bigskip
\bigskip
\section{Reduced-order Surrogate}
\label{sec:ros}

In this section, we focus our attention on constructing a surrogate that captures the dependence of
uncertainty in NEMD predictions of the bulk thermal conductivity ($\kappa$) on input uncertainty
in the SW potential parameters. The surrogate is a powerful tool that greatly minimizes
the computational
effort required for forward propagation of the uncertainty from input parameters to the output,
Sobol\textquotesingle~sensitivity analysis, and Bayesian calibration of the uncertain model parameters. Once again,
we use polynomial chaos (used to construct response surfaces for discrepancy in
Section~\ref{sec:response}) to construct the surrogate of the following functional form:

\be
\kappa  = \sum\limits_{\bm{s}\in\mathcal{A}} c_{\bm{s}}(T)\Psi_{\bm{s}}(\bm{\xi})
\ee

\begin{figure}[htbp]
 \begin{center}
  \includegraphics[width=0.65\textwidth]{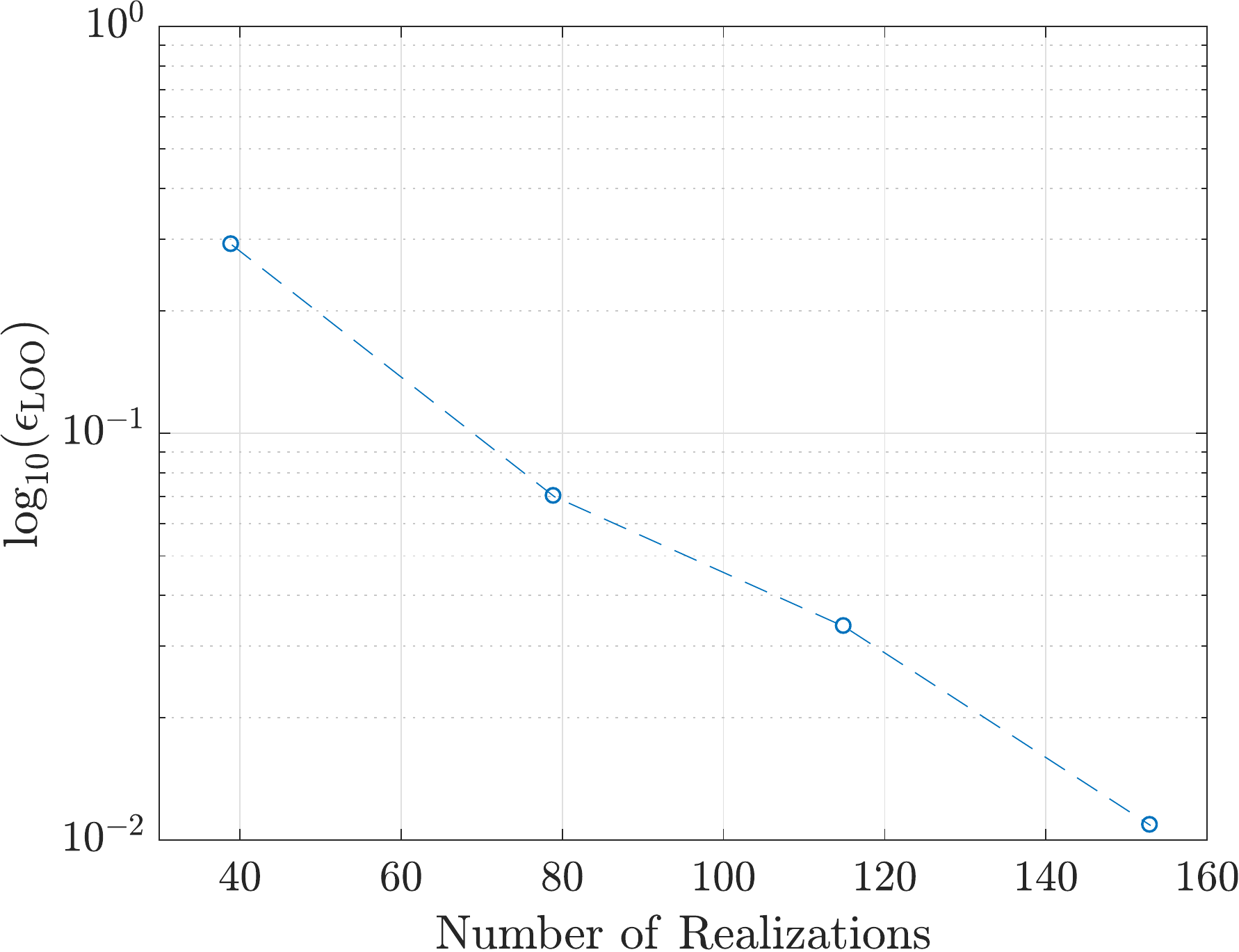}
\caption{A convergence study for the 5D PCE wherein the leave-one-out
error, $\epsilon_{\tiny{\mbox{LOO}}}$ is plotted against the number of
realizations or NEMD runs used to estimate the PC coefficients.}
\label{fig:loo}
\end{center}
\end{figure}

As discussed earlier in Section~\ref{sec:response}, several strategies are available to
estimate the PC coefficients, $c_{\bm{s}}$. However, since the polynomial basis functions
($\Psi_{\bm{s}}(\bm{\xi}$)) are
relatively high dimensional, we use a computationally efficient approach proposed by
Blatman and Sudret to construct a PCE with sparse basis($\mathcal{A}$) using the LAR
algorithm~\cite{Blatman:2011}. Furthermore, since the NEMD simulations are compute-intensive,
estimating the PC coefficients in the 7D parameter space would still require a large amount of
computational resources. Hence, we explore the possibility of reducing the dimensionality of
the surrogate. For this purpose, we 
exploit our observations in Figure~\ref{fig:ub} where a significant
jump in the $\hat{\mathcal{C}_i\mu_i}$ estimate is seen from $A$ to $B$ and thereby construct the 
PC surrogate in a 5D parameter space by fixing $B$ and $p$ at their nominal values. In the above
equation, $\bm{\xi}:~\{\xi_1(A),\xi_2(q),\xi_3(\alpha),\xi_4(\lambda),\xi_5(\gamma)\}$ is a set
of five canonical random variables, $\xi_i$ distributed uniformly in the interval [-1,1].
Prior intervals for the uncertain SW parameters are considered to be
$\pm~10\%$ of their respective nominal estimates except for $q$ in which case it is [0,0.1]. In
Figure~\ref{fig:loo}, we plot the leave-one-out cross-validation error 
($\epsilon_{\tiny{\mbox{LOO}}}$)~\cite{Blatman:2010}, defined
below in Eq.~\ref{eq:loo}, against the number of model realizations used to construct the 5D
PC surrogate. The software, UQLab~\cite{Marelli:2014} was used for estimating $\epsilon_{\tiny{\mbox{LOO}}}$
and constructing the surrogate. The leave-one-out cross validation error ($\epsilon_{\tiny{\mbox{LOO}}}$) is
computed as:

\be
\epsilon_{\tiny{\mbox{LOO}}} = \frac{\sum\limits_{i=1}^{N}\left(\mathcal{M}(\bm{x}^{(i)}) - 
\mathcal{M}^{PCE\setminus i}(\bm{x}^{(i)})\right)^{2}}{\sum\limits_{i=1}^{N}
\left(\mathcal{M}(\bm{x}^{(i)}) - \hat{\mu}_Y\right)^2}
\label{eq:loo}
\ee 

\noindent where $N$ denotes the number of realizations, $\mathcal{M}(\bm{x}^{(i)})$ is the
model realization and $\mathcal{M}^{PCE\setminus i}(\bm{x}^{(i)})$ is the corresponding PCE estimate
at $\bm{x}^{(i)}$. Note that the PCE is constructed using all points except $\bm{x}^{(i)}$.
The quantity, $\hat{\mu}_Y$ = $\frac{1}{N}\sum\limits_{i=1}^{N}\mathcal{M}(\bm{x}^{(i)})$
is the sample mean of the realizations.

It is found that in order for the PCE to converge to an accuracy of~$\mathcal{O}(10^{-2})$
with respect to $\epsilon_{\tiny{\mbox{LOO}}}$, we require approximately 160 NEMD runs. In the
following section, we focus on verifying the accuracy of the 5D PCE against the set of available
NEMD predictions in the original 7D parameter space. 

\subsection{PC Surrogate Verification}

The accuracy of the 5D PC surrogate is verified using two different strategies. The first strategy involves
computing the relative L-2 norm of the difference between the available NEMD predictions (used earlier
to estimate DGSM in Section~\ref{sec:sense}) and estimates using the 5D surrogate as follows:

\be
\epsilon_{\mbox{\tiny{L-2}}} = 
\frac{\left[\sum\limits_{i=1}^{N=25}\left(\mathcal{M}(\bm{\theta}_{\mbox{\tiny{7D}}}^{(i)}) - 
\mathcal{M}^{PCE}(\bm{\theta}_{\mbox{\tiny{5D}}}^{(i)})\right)^{2}\right]^{\frac{1}{2}}}{\left[\sum_{i=1}^{N}
\left(\mathcal{M}(\bm{\theta}_{\mbox{\tiny{7D}}}^{(i)})\right)^2\right]^{\frac{1}{2}}} \approx 6.88\times 10^{-2}
\ee
 
\noindent where $\mathcal{M}(\bm{\theta}_{\mbox{\tiny{7D}}}^{(i)})$ is the NEMD prediction in the original 7D
parameter space, and $\mathcal{M}^{PCE}(\bm{\theta}_{\mbox{\tiny{5D}}}^{(i)})$ is the corresponding 
estimate using the reduced order surrogate (5D). 
Since $\epsilon_{\mbox{\tiny{L-2}}}$ is found to be $\mathcal{O}(10^{-2})$, the 5D surrogate can be considered as
reasonably accurate from the perspective of relative L-2 error norm. 

\begin{figure}[htbp]
 \begin{center}
  \includegraphics[width=0.70\textwidth]{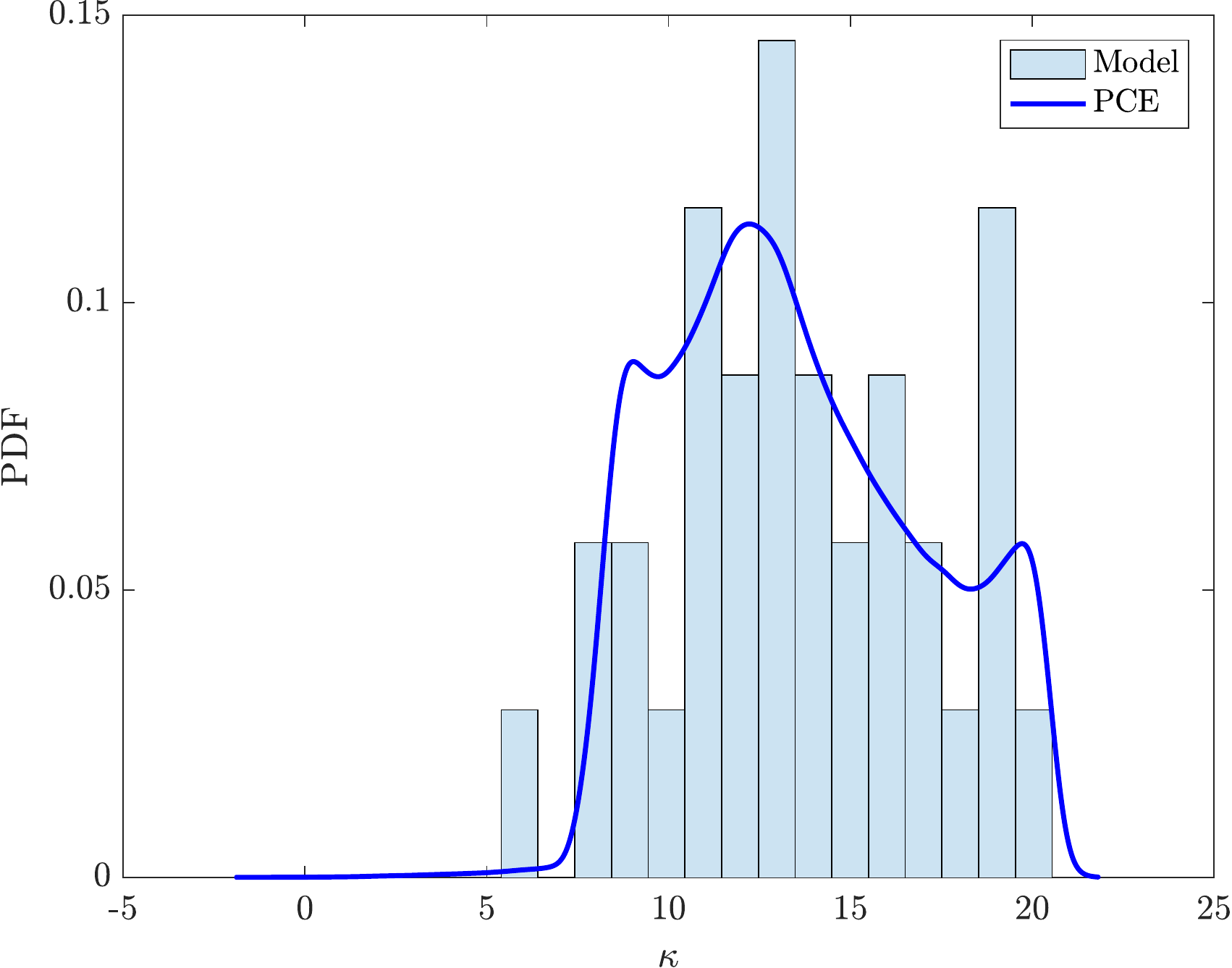}
\caption{Comparison of bulk thermal conductivity ($\kappa$) distribution of Si based on a histogram plot
using NEMD predictions (Model) at 25 points in the 7D parameter space and a probability distribution obtained 
using kernel density estimation of  reduced-order surrogate estimates of $\kappa$ for 10$^6$ samples in the 5D
parameter space.}
\label{fig:verify}
\end{center}
\end{figure}

In the second strategy, we compare NEMD
predictions and estimates from the 5D surrogate in a probabilistic sense. As shown in Figure~\ref{fig:verify}, a 
histogram plot based on the available set of NEMD predictions for bulk thermal conductivity in the 7D parameter 
space is compared with its probability distribution, obtained using the reduced order surrogate and
10$^6$ samples in the 5D parameter space described by the SW potential parameters:~$\{A,q,\alpha,\lambda,\gamma\}$. It is observed that the probability density function (PDF) based on estimates from the reduced-order surrogate compares
favorably with the histogram. Specifically, the corresponding mode values for $\kappa$ from the two plots are
in close agreement, and the PDF reasonably captures the peaks as well as the spread in
the bulk thermal conductivity distribution as observed in the histogram. 
Hence, the reduced-order surrogate is verified for accuracy in both cases. 
\bigskip

\begin{figure}[htbp]
 \begin{center}
  \includegraphics[width=0.65\textwidth]{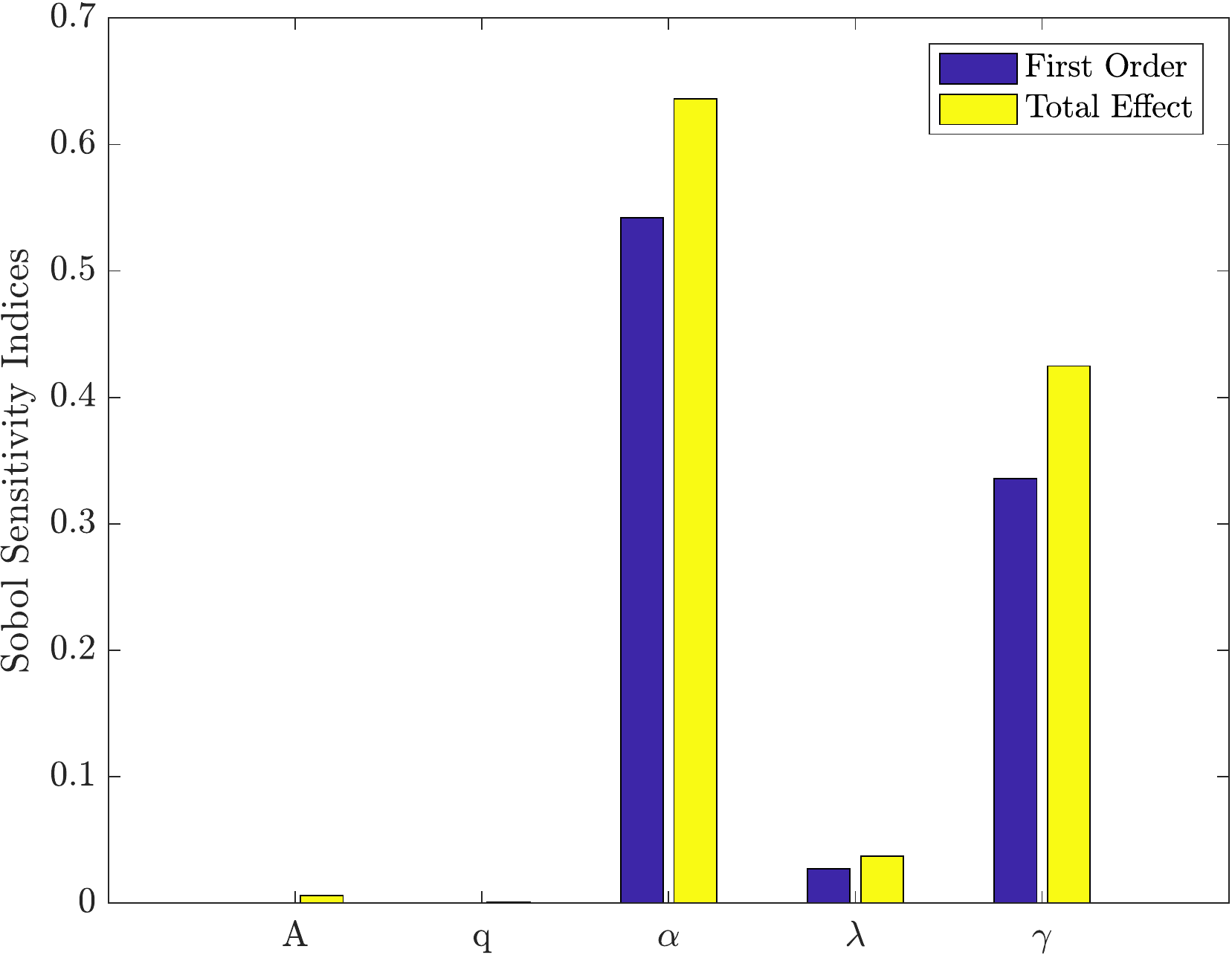}
\caption{Sobol\textquotesingle~first-order and total effect sensitivity indices of the SW potential parameters
w.r.t the bulk thermal
conductivity as obtained using the reduced order PC surrogate and 10$^{6}$ samples in the 5D parameter space. }
\label{fig:gsa}
\end{center}
\end{figure}

As mentioned earlier, a PC surrogate can be used to estimate the Sobol\textquotesingle~global sensitivity indices in a 
straightforward manner~\cite{Sudret:2008}. The Sobol\textquotesingle~first order and total effect sensitivity indices estimated
using the reduced-order surrogate and $10^{6}$ samples in the 5D parameter space are plotted using bar-graphs in
Figure~\ref{fig:gsa}. 

It is found that $\kappa$ is predominantly sensitive towards the choice of $\alpha$ and
$\gamma$. This observation is consistent with our initial findings based on DGSM using 25 samples in the 7D
parameter space. However, the DGSM estimate for $\gamma$ was found to be the highest in that case. 
While sensitivity towards $A$, $q$, and $\lambda$ is observed to be comparatively less in both cases, the 
Sobol\textquotesingle~indices for the three parameters are estimated to be smaller by an order of magnitude
 compared to those
for $\alpha$ and $\gamma$. Large quantitative disagreement in parametric sensitivity between DGSM and the 
Sobol\textquotesingle~indices is however not unexpected essentially because the two metrics differ by construction.
The former is based
on an expectation of partial derivatives while the later is based on variance. Nevertheless, significant qualitative
agreement pertaining to parameter importance in the two approaches is quite encouraging. 
Moreover, verification of the reduced-order
surrogate for accuracy increases our confidence in implementing DGSM to ascertain the relative importance of the
SW potential parameters and hence perform uncertainty analysis for the present application with minimal
computational effort.

\bigskip
\bigskip
\section{Bayesian Calibration}
\label{sec:bayes}

As discussed earlier in Section~\ref{sec:intro}, the underlying methodology for determining the nominal 
estimates for SW potential parameters did not account for measurement error, inadequate functional
form of the potential, inherent noise in MD predictions, and parametric uncertainties. Hence, there is
a possibility of improving the estimates since using the same set of values for a wide variety of systems
and applications is not ideal. A robust approach to calibrating the parameters in the presence of
such uncertainties is made possible by using a Bayesian framework. The methodology aims at evaluating
the so-called joint posterior probability distribution (referred to as the `posterior') of the uncertain model 
parameters to be calibrated, using Bayes' rule:

\be
\mathcal{P}(\bm{X}\vert \bm{Y}) \propto \mathcal{P}(\bm{Y}\vert\bm{X})\mathcal{P}(\bm{X})
\ee

\noindent where $\bm{X}$ is the set of uncertain model parameters, and $\bm{Y}$ is the available set of
experimental data i.e. bulk thermal conductivity at different temperatures in the present case. 
We exploit our findings based on sensitivity analysis in
Figure~\ref{fig:gsa} and focus on calibrating $\alpha$ and $\gamma$ i.e. $\bm{X}:\{\alpha,\gamma\}$.
$\mathcal{P}(\bm{X}\vert \bm{Y})$ is regarded as the posterior, $\mathcal{P}(\bm{Y}\vert\bm{X})$ is the
`likelihood', and $\mathcal{P}(\bm{X})$ is the joint prior probability distribution (referred to as the `prior') of $\bm{X}$.
The likelihood accounts for measurement error, and the discrepancy between experiments and model
predictions, whereas, the prior is an initial guess for the distribution of uncertain model parameters in an
interval. It also accounts for the availability of the expert opinion pertaining to their estimates. 
The posterior provides an estimate of the most likely value of the uncertain model parameters based on
prior uncertainty, experimental data used for calibration and the associated measurement error, and model
discrepancy. Additionally, the posterior is often used to quantify the uncertainty associated with model predictions. Several algorithms based on the Markov chain Monte Carlo (MCMC) technique are available for  
sampling the posterior~\cite{Haario:2001, Haario:2006,Xu:2014}.

\begin{figure}[htbp]
 \begin{center}
  \includegraphics[width=0.70\textwidth]{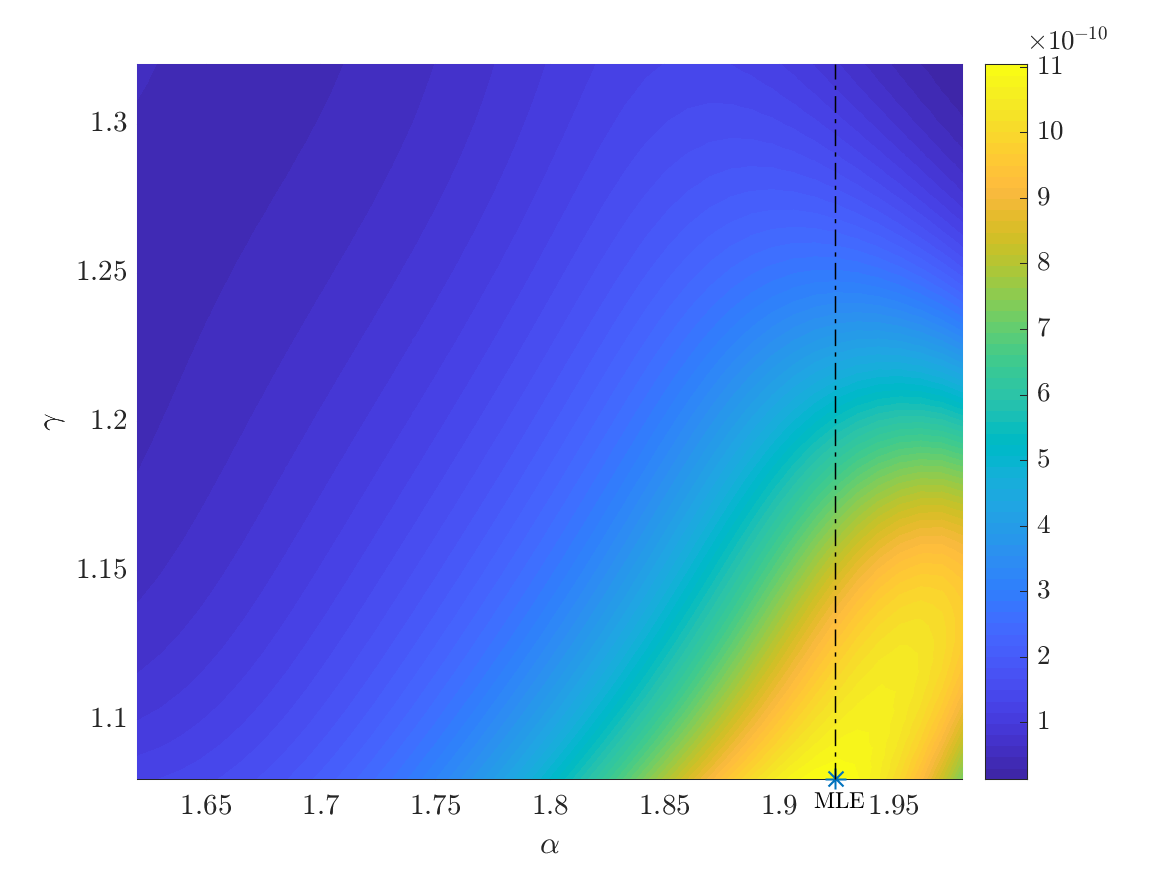}
  \\ (a)
  \begin{tabular}{cc}
  \includegraphics[width=0.50\textwidth]{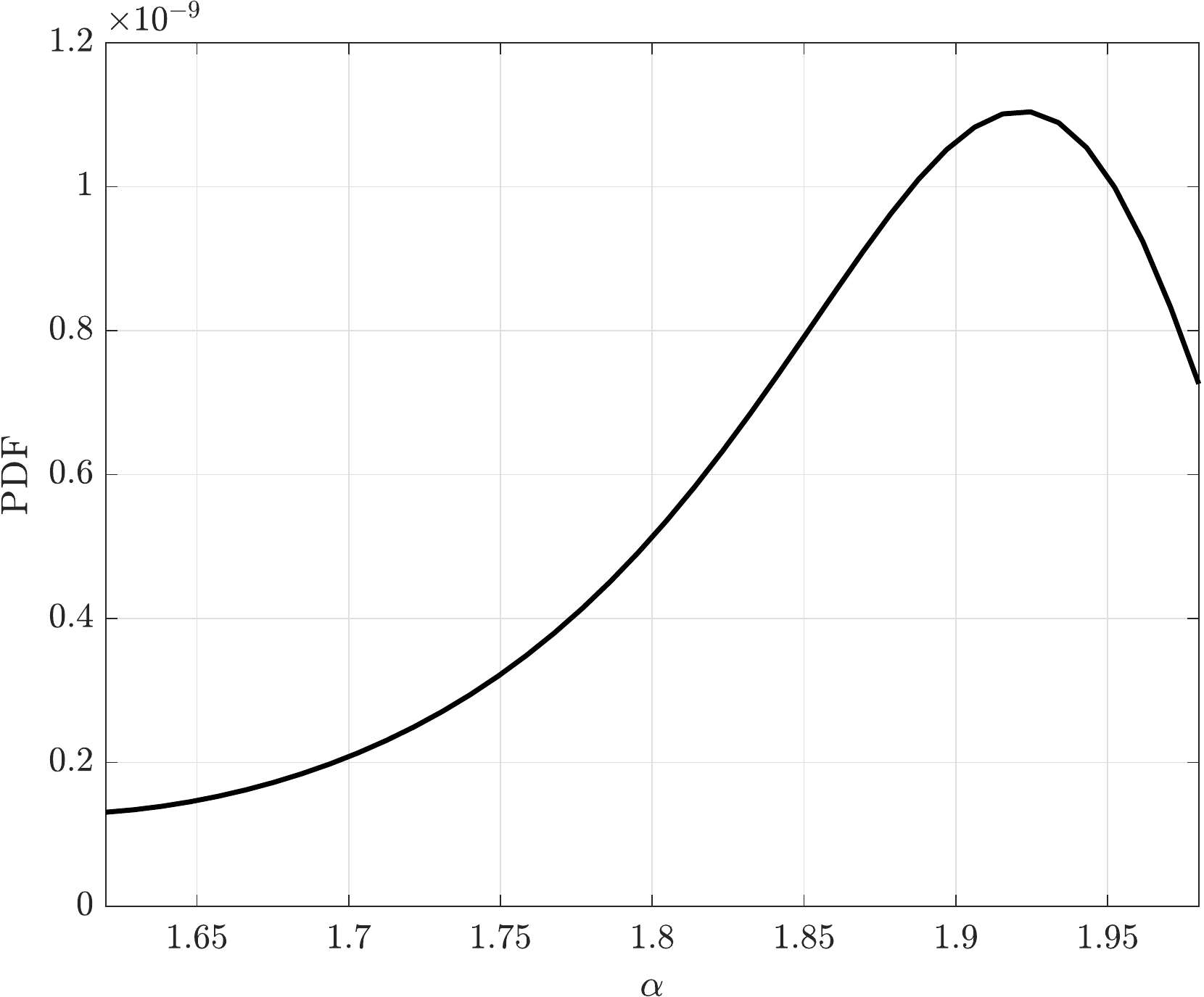}
  &
  %\hspace{3mm}
  \includegraphics[width=0.50\textwidth]{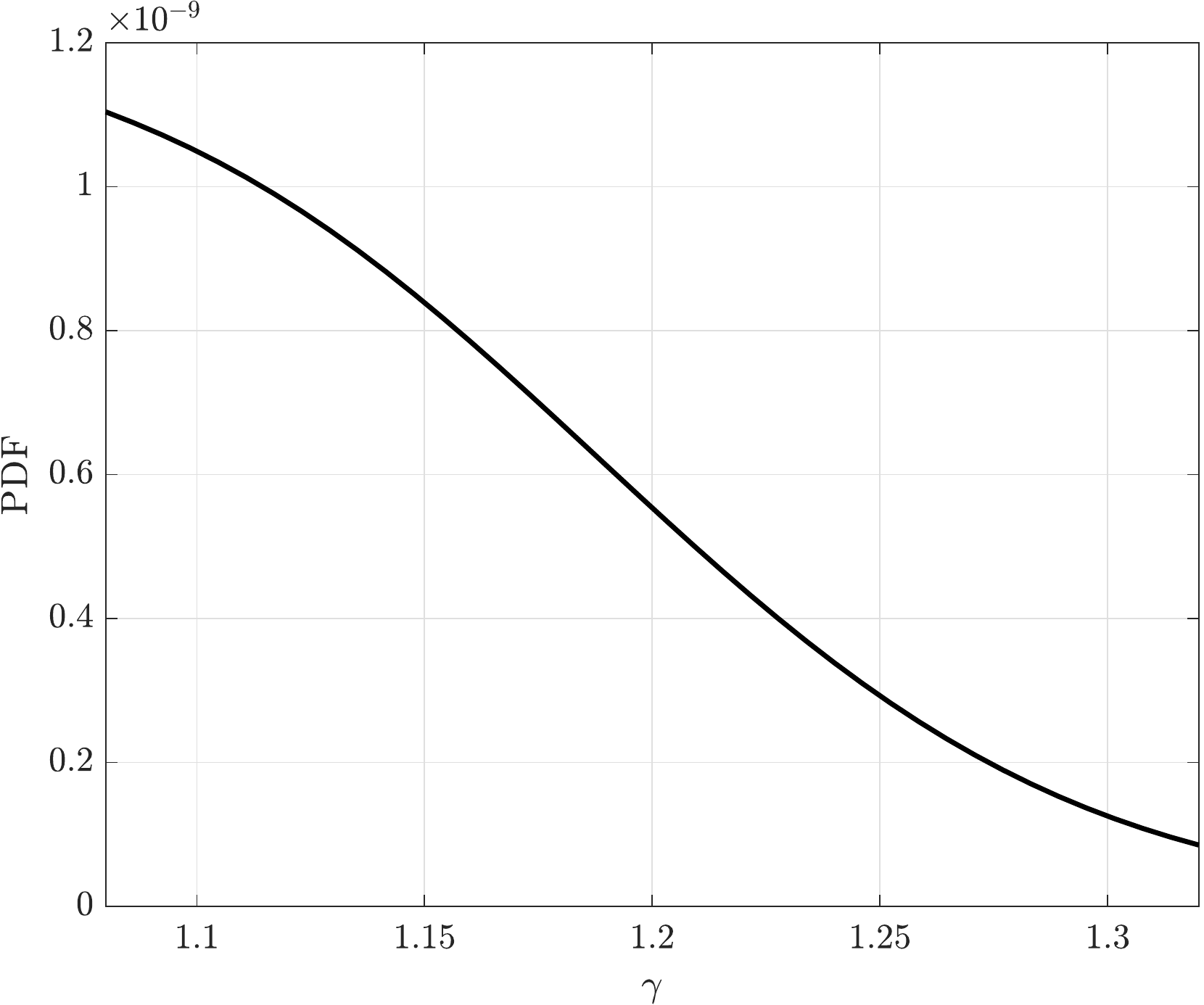}
  \\ (b) & (c)
  \end{tabular}
\caption{(a) The joint likelihood of ($\alpha$,$\gamma$) as estimated using Eq.~\ref{eq:like} is plotted
on a 2D cartesian grid. The maximum likelihood estimate (MLE) is also highlighted.
Marginal likelihoods for $\alpha$ and $\gamma$ are plotted in (b) and (c) respectively.}
\label{fig:like}
\end{center}
\end{figure}

Evaluating the joint posterior of $\alpha$ and $\gamma$ using MCMC typically requires a large amount of
computational effort and is not the focus of this work. Instead, we compute and plot the joint likelihood on
a 2D cartesian grid described by $\alpha$ and $\gamma$ using Eq.~\ref{eq:like} as illustrated in
Figure~\ref{fig:like}(a). For this purpose, we consider the priors of $\alpha$ and $\gamma$ to be
independent and uniformly distributed in the intervals, [1.62,1.98] and [1.08,1.32] respectively. 
Consequently, the posterior is proportional to the likelihood, considered to be a Gaussian:

\be
\mathcal{P}(\bm{Y}\vert\bm{X}) = \frac{1}{\sqrt{2\pi\sigma^2}}\exp\left[-\frac{(\kappa_{\tiny{\mbox{E}}} - 
\kappa_{\tiny{\mbox{MD}}})^2}{2\sigma^2}\right]
\label{eq:like}
\ee

\noindent where $\sigma$ is the standard deviation of the measurement error, and
$(\kappa_{\tiny{\mbox{E}}} - \kappa_{\tiny{\mbox{MD}}})$ is the discrepancy between 
NEMD predictions~($\kappa_{\tiny{\mbox{MD}}}$)
and experimental data~($\kappa_{\tiny{\mbox{E}}}$). Experimental data for $\kappa_{\tiny{\mbox{E}}}$ at 300~K
(149~W/m/K~\cite{Shanks:1963}) is used to compute the joint likelihood. It must be noted that the likelihood
function in Eq.~\ref{eq:like} could be refined further by accounting for a model discrepancy term and hence
calibrating the associated parameters in addition to the uncertain model inputs~\cite{Kennedy:2001,Ling:2014}. 

Marginal distributions for
$\alpha$ and $\gamma$ are shown in Figure~\ref{fig:like}(b) and Figure~\ref{fig:like}(c) respectively. 
As mentioned above, the joint likelihood plot is based on the bulk thermal conductivity measurement at 300~K.
An enhanced set of experimental measurements at different bulk temperatures would help improve the
accuracy of the calibration process considering the measurement noise is not too large. Moreover, it would
help capture the correlation between calibration parameters across temperatures at which the data is
available. However, a reduced-order surrogate would be needed at each temperature in order to make the
MCMC tractable.

\bigskip
\bigskip
\section{Summary and Discussion}
\label{sec:disc}

In this paper, we have attempted to identify and address some of the challenges
pertaining to uncertainty quantification of bulk thermal conductivity predictions 
using non-equilibrium molecular dynamics (NEMD) simulations. Specifically, we focused
on investigating the impact of system size, and fluctuations in the applied thermal
gradient on predictions. In order to quantify the discrepancy between NEMD
predictions and experiments, response surfaces were  
constructed at bulk temperatures, $T$ = 300~K, 500~K, and 1000~K.  
It was found that the discrepancy is predominantly impacted by size while the 
effect of fluctuations in the applied thermal gradient is negligible in the considered
interval. The response surface approach presented here relies on a small number of
MD runs and enables an accurate estimation of discrepancy at a given temperature
and a point in the 2D parameter space described by system-size and the applied
thermal gradient. 

A possible enhancement of nominal SW parameter estimates for a given application
and the choice of material system is also highlighted in this work. To enable this,
we focus our efforts on understanding the sensitivity of predictions on individual potential
parameters. In order to reduce the computational effort, we estimate the derivative-based
sensitivity measures (DGSM) and hence the upper bound
on Sobol\textquotesingle~total-effect sensitivity index using random samples in the 7D parameter space. 
While individual measures for the 7 parameters are not too distant from each other, the predictions
seem to be most sensitive towards $\gamma$. Sensitivity measures for $\alpha$,
$\lambda$, $q$, and $A$ are found to be comparable while those for $B$ and $p$
are relatively small. 

A polynomial chaos surrogate model for the bulk thermal conductivity (observable)
as a function of SW potential
parameters at the bulk temperature of 300~K is constructed. The surrogate helps reduce the
computational effort required for forward propagation of the uncertainty from
parameters to the observable as well as for estimating the Sobol\textquotesingle~sensitivity indices.
Furthermore, the surrogate could be used to accelerate parameter calibration in a
Bayesian setting. However, since the surrogate relies on NEMD predictions, the
underlying computational effort is nevertheless significantly large. To circumvent
this challenge, we exploit our initial findings based on DGSM and construct
a reduced-order surrogate in 5 dimensions by fixing the parameters, $B$ and $p$.
We verify its accuracy by estimating the relative
L-2 norm of the error between NEMD predictions in the full space and the reduced-order
surrogate predictions, and by comparing the probability density of the
bulk thermal conductivity in Figure~\ref{fig:verify}. Furthermore, our initial
sensitivity trends based on DGSM-analysis using 25 samples seem to agree favorably
with Sobol\textquotesingle~sensitivity analysis based on 10$^{6}$ samples. Hence, it can be said
that DGSM-based analysis with a few samples could offer huge computational gains
by reliably reducing the dimensionality of a surrogate for uncertainty analysis. 

Finally, we highlight key aspects of parameter calibration in a Bayesian setting.
The underlying motivation stems from the fact that the nominal estimates of the SW
parameters did not consider measurement error, simulation noise, model form error,
and parametric uncertainties. Calibration in a Bayesian framework allows us to
incorporate such errors and uncertainties in an efficient manner and provides
a joint posterior distribution of the uncertain parameters. To minimize computational
costs pertaining to the calibration process, we suggest the following sequence of
steps: First, perform DGSM analysis to 
identify parameters that are not so important. Second, construct a reduced-order
surrogate based on DGSM-analysis and verify its accuracy. Third, compute the
Sobol\textquotesingle~sensitivity indices using the surrogate to identify parameters for calibration. 
Fourth, construct a second surrogate in the reduced subspace described by the
calibration parameters, and quantify the surrogate error to be accounted for
in calibration. Fifth, evaluate the joint posterior using an efficient MCMC-based algorithm
such as adaptive Metropolis~\cite{Haario:2001} and its variants~\cite{Haario:2006, Green:2001}.

It is important to note that the strategies presented 
in this work are not restricted to a Si bar and the Stillinger-Weber potential, and
could be extended to a wide range of applications and inter-atomic potentials for
the purpose of uncertainty quantification.

\bigskip
\bigskip
\section*{Acknowledgment}
\label{sec:ack}

M. Vohra and S. Mahadevan gratefully acknowledge funding support from the
National Science Foundation (Grant No. 1404823, CDSE Program). Molecular
dynamics simulations were performed using resources at the advanced computing
center for research and education (ACCRE) at Vanderbilt University. 
M. Vohra is grateful to the ACCRE staff for their technical guidance and support.
This work partially used the Extreme Science and Engineering Discovery Environment 
(XSEDE), which is supported by National Science Foundation grant number ACI-1053575.
The computational resources were awarded through XSEDE research project, TG-CTS170020.
M. Vohra would also like to sincerely thank Dr. Alen Alexanderian at NC State 
University for insightful discussions pertaining to the derivative-based sensitivity analysis
in this work.

\bigskip
\bigskip
\bibliographystyle{unsrt}
\bibliography{REFER}
%
%\clearpage
%
%\input{figures}
%\clearpage

\end{document}